\documentclass[journal]{IEEEtran}
\usepackage{latexsym,epsf,times,amsmath,color,amssymb,indentfirst,fancyhdr,colortbl,
bm}
\usepackage{subcaption}
 \usepackage{graphicx}
  \usepackage{cite}
\newcommand{\nc}{\newcommand}
\nc{\be}{\begin{equation}}
\nc{\ee}{\end{equation}}
\def\calN{{\mathcal{N}}}

\nc{\Abf}{\mathbf{A}}
\nc{\Cbf}{\mathbf{C}}
\nc{\Dbf}{\mathbf{D}}
\nc{\Fbf}{\mathbf{F}}
\nc{\Gbf}{\mathbf{G}}
\nc{\Hbf}{\mathbf{H}}
\nc{\Ibf}{\mathbf{I}}
\nc{\Lbf}{\mathbf{L}}
\nc{\Nbf}{\mathbf{N}}
\nc{\Pbf}{\mathbf{P}}
\nc{\Qbf}{\mathbf{Q}}
\nc{\Rbf}{\mathbf{R}}
\nc{\Sbf}{\mathbf{S}}
\nc{\Ubf}{\mathbf{U}}
\nc{\Vbf}{\mathbf{V}}
\nc{\Wbf}{\mathbf{W}}
\nc{\Xbf}{\mathbf{X}}
\nc{\Ybf}{\mathbf{Y}}
\nc{\Zbf}{\mathbf{Z}}
\nc{\abf}{\mathbf{a}}
\nc{\bbf}{\mathbf{b}}
\nc{\hbf}{\mathbf{h}}
\nc{\gbf}{\mathbf{g}}
\nc{\nbf}{\mathbf{n}}
\nc{\pbf}{\mathbf{p}}
\nc{\qbf}{\mathbf{q}}
\nc{\rbf}{\mathbf{r}}
\nc{\sbf}{\mathbf{s}}
\nc{\ubf}{\mathbf{u}}
\nc{\vbf}{\mathbf{v}}
\nc{\wbf}{\mathbf{w}}
\nc{\xbf}{\mathbf{x}}
\nc{\ybf}{\mathbf{y}}
\nc{\zbf}{\mathbf{z}}
\nc{\unobf}{\mathbf{1}}
\nc{\zerobf}{\mathbf{0}}
\nc{\vbfs}{\mathbf{\scriptsize v}}
\nc{\Phibf}{\mathbf{\Phi}}
\nc{\Psibf}{\mathbf{\Psi}}
\nc{\Thetabf}{\mathbf{\Theta}}
\nc{\Lambdabf}{\mathbf{\Lambda}}
\nc{\Sigmabf}{\mathbf{\Sigma}}
\nc{\Omegabf}{\mathbf{\Omega}}
\nc{\xibf}{{\mbox{\boldmath $\xi$}}}
\nc{\betabf}{{\mbox{\boldmath $\beta$}}}
\nc{\rhobf}{{\mbox{\boldmath $\rho$}}}
\nc{\nubf}{{\mbox{\boldmath $\nu$}}}
\nc{\xibfs}{{\mbox{\boldmath \scriptsize $\xi$}}}
\nc{\diag}{\text{diag}}
\nc{\sign}{\text{sign}}
\nc{\E}{\text{E}}
\def\defeq{\stackrel {\scriptscriptstyle \Delta}{=}}

\IEEEoverridecommandlockouts
\makeatletter
\title{Improved SAR Imaging \\ Via Cross-Learning from Camera Images}
\author{\IEEEauthorblockN{Shahzad Gishkori, David Wright, Liam Daniel, Marina Gashinova and Bernard Mulgrew}
\thanks{S. Gishkori, D. Wright and B. Mulgrew are with Institute for Digital Communications (IDCOM), The School of Engineering, The University of Edinburgh, United Kingdom. 
Emails: \{s.gishkori, d.wright, bernie.mulgrew\}@ed.ac.uk}
\thanks{L. Daniel and M. Gashinova are with Microwave Integrated System Laboratory (MISL), School of Electronic, Electrical and Systems Engineering, University of Birmingham, United Kingdom. 
Emails: \{l.y.daniel, m.s.gashinova\}@bham.ac.uk}
\thanks{This work was supported by Jaguar Land Rover and the UK-EPSRC grants EP/N012240/1 \& EP/N012372/1 as part of the jointly funded Towards Autonomy: Smart and Connected Control (TASCC) Programme.}
}
\providecommand{\keywords}[1]{\textbf{\textit{Index terms---}} #1}
\begin{document}
\maketitle
%------------------------------------------------------------------------------------------------------------------
\vspace{-0.3in}
\begin{abstract}
\noindent
In this paper, we propose a novel concept of cross-learning, in order to improve SAR images by learning from the camera images.
We use a multi-level abstraction approach to materialise knowledge transfer between the two modalities.
We also compare the performance of other possible approaches. 
We provide experimental results on real data to validate the proposed concept.
\end{abstract}
%------------------------------------------------------------------------------------------------------------------
\section{Introduction}
\label{sect:intro}

Synthetic aperture radar (SAR) \cite{Carrara_95,Jakowatz_SpotSAR_1996,Soumekh_SAR_SP_99} can provide high-resolution images. 
Substantial amount of work is available to enhance SAR image quality in terms of denoising and/or super-resolution (see e.g., \cite{Cetin_SAR_2014,SG_GSP_2019} and references therein).
However, SAR image quality is still not at par with that of optical sensors, e.g, camera and lidar.
Nonetheless, due to its ability to generate images, even in adverse weather conditions, SAR is emerging as a new imaging mode for automotive scenarios \cite{SG_FSSAR_2018,SG_FSSARMOV_2019,Bilik_2019}.
However, most of the previous work focuses on improving the SAR image by assuming SAR to be a stand-alone sensor, without any interaction with any sensor of a different modality. In automotive (especially, autonomous driving) scenarios, a car may be equipped with multiple sensors, e.g, radar, lidar, camera, etc \cite{Google_car_2011,MultiSensorFusion_2014}. 
Therefore, it is natural to explore if a SAR image can be improved by using images from other sensors of different modalities, i.e., exploiting the framework of multi-modal fusion \cite{Khaleghi_Multisensor_2013,Adali_MultimodalFusion_2015,DeepMultimodal_2017}.
This motivation also forms the basis for our present paper.

Multi-modal fusion is a very generic concept, combining data/information from diverse modalities in order to enhance the achievement of a common objective, e.g., creating a unified sensing system, improving decision making, identifying/extracting specific features, etc. 
The key property is diversity, i.e., multiple modalities complementing each other in achieving a common goal in a way that cannot be achieved with a single modality \cite{Adali_MultimodalFusion_2015}.
Despite the unquestionable motivation for multi-modal fusion, the real challenge is how to exactly exploit this diversity. The reasons are that different modalities may be driven by different underlying variables or they may operate on different physical principles, etc. Therefore, finding a direct correspondence/correlation between them might not be straightforward. Some effort has been expended to devise a certain level of abstraction, e.g., \cite{Khaleghi_Multisensor_2013,Structured_Fusion_2015} (and references therein). However, more research needs to be done.

SAR and lidar are active sensor modalities.
Some work on the fusion of hyperspectral SAR images and lidar images, in remote sensing domain, has appeared recently, e.g, \cite{Hyperspectral_LiDAR_2014,Dalla_Multimodality_Challenge_2015}.
A camera, in contrast, is a passive sensor modality and exploits illumination from other sources. %(operates on natural illumination.) 
Its ranging estimates (e.g., obtained by using the depth-maps for a stereo camera) are not as good as that of radar or lidar. However, it can provide very good image resolution. 
Due to the very different dynamics of SAR and camera sensors, e.g., operating principles, coordinate systems, pixel resolution, etc, it is quite hard to register and fuse their respective images.  Generally, in the available work, e.g., \cite{MultiSensorFusion_2014}, radar is primarily used as a detection sensor instead of an imaging sensor.
Thus, according to our knowledge, not much work is available on the fusion of SAR and camera images. 

In this paper, we focus on fusing SAR and camera images at the data level without (strictly) registering the respective images of the two sensors. 
To emulate automotive scenarios, we basically consider short-range radar sensing for extended targets (instead of point scatterers). 
Our primary aim is to reconstruct high-resolution SAR images, i.e, reconstructing the physical details of the extended target. In order to do this, we learn certain features of the target from camera images. Since these features have been learnt from a very different modality, we name this process as cross-learning. 
Note, cross-learning may have some overlap with transfer-learning \cite{Survey_Transfer_Learning_2010}. However, the emphasis in the former is on different modalities.
Cross-learning (from camera images) to improve SAR imaging is a very new concept and it has the potential to become a new area of research given the amount of challenges and opportunities associated with it.
In this paper, we present an approach to materialise this concept.
Our basic premise is the observation that despite difference in resolution and viewing perspective, both modalities try to capture the same physical geometry of the target. Therefore, correspondence or correlation between the two sensors does exist in some latent- or intrinsic-dimensional representation of their respective images.
This may potentially circumvent the need for strict inter-sensor mapping/registration.

Traditionally, manifold learning techniques (linear or non-linear) \cite{bishop,Survey_Manifold_2008,VanDerMaaten_Manifold} have been used to retrieve low-dimensional representation of a high-dimensional data for a wide range of tasks, e.g., 
detection, estimation, classification, visualisation, fusion, etc \cite{Roweis_00,Belkin_2003,manifold_05,Coifman_2006,Baraniuk_Manifold_2010}.
However, most of these tasks are (best) carried out in the manifold domain without the aim of reconstructing the high-dimensional data.
In our case, we need to, {\it i)} create the respective manifolds of SAR and camera images to generate the intrinsic-dimensional or manifold-domain representation, 
{\it ii)} learn extra features from camera manifold and transfer it to the SAR manifold,
{\it iii)} reconstruct the SAR image in its high-dimensional or image-domain representation.
Since, we do the learning in the manifold-domain and then reconstruct the original image-domain, we can only use linear manifolds, e.g, principal component analysis (PCA). Non-linear manifolds, e.g., Laplacian eigenmaps (LE), locally linear embedding (LLE), Hessian eigenmaps, diffusion maps etc, provide efficient low-dimensional representation. However, they cannot be transformed/projected back to the image domain. 
Manifold alignment \cite{Coifman_2006,Ham_ManifoldAlign_2005,Wang_ManifoldAlign_2008} has been an effective way of transferring knowledge/information between different datasets. 
Similarly, in the case of super-resolution of face images, building on a two step approach of global and then local features adjustment \cite{Liu_FaceHallucinating_2007,Chang_SuperRes_LLE_2004}, some authors, e.g., \cite{Huang_SuperResFace_2010,Huang_SuperResFace_2011} have advocated the creation of a coherent subspace over the manifolds for efficient transfer of knowledge. Due to an extra layer of abstraction, the latter approach has the ability to work with a modest amount of training samples as well as to compensate for choosing a linear manifold instead of a non-linear manifold (if required).
Now, in case of transferring information from a camera image to a radar image, i.e., cross-learning, there are multiple challenges, e.g., the modalities are different, coordinate systems are different therefore the two sensors cannot be fully registered with each other, there is substantial disparity in resolution, choice of manifolds is limited due to reconstruction requirement, etc. Thus, in order to circumvent these challenges, a multi-stage abstraction may be the right course of action. 
To this end, we follow the approach of \cite{Huang_SuperResFace_2010}. 
We create PCA-based manifolds for both the sensors and generate a coherent subspace by using the canonical correlation analysis (CCA) \cite{Hardoon_CCA_2004}. Then, we use LLE \cite{Roweis_00} to learn/adjust the neighbourhood embedding of the coherent subspace from the camera to the radar, followed by recovering the improved SAR image. 
Note, the input SAR images are generated by our recently proposed forward-scanning SAR (FS-SAR) \cite{SG_FSSAR_2018} mode for the automotive scenarios, albeit, the synthetic aperture considered here is circular instead of linear, i.e., a circular-scanning SAR (CiS-SAR).  
\\
\noindent
{\bf \em Contributions}.
The following are the main contributions of this paper.
\begin{itemize}
\item 
We present a novel concept of improving SAR images via cross-learning from camera images.
\item
We show that multiple levels of abstraction can help circumvent the challenges of knowledge transfer in these different modalities.
\item
We consider a CiS-SAR mode generated SAR images as input to the cross-learning frame-work.
\item
We present qualitative performance results based on real-data obtained in our lab controlled experimental setup. 
\end{itemize}
\noindent
{\bf \em Organisation}.
Section \ref{sect:sig_mod} provides the system model, Section \ref{sect:xlearn} elaborates on the realisation of cross-learning via a multi-level abstraction approach, Section \ref{sec:sim} provides experimental results and performance comparisons, and Section \ref{sec:concl} gives the conclusions.
\\
\noindent
{\bf \em Notations}.
Matrices are in upper case bold while column vectors are in lower case bold,
$(\cdot)^T$ denotes transpose,
$[\abf]_{i}$ is the $i$th element of $\abf$
and 
$[\Abf]_{i,j}$ is the $ij$th element of $\Abf$,
$\hat{\abf}$ is the estimate of $\abf$,
$\defeq$ defines an entity,
and 
the $\ell_p$-norm is denoted as $||\abf||_p = (\sum_{i=1}^{N} |[\abf]_{i}|
^p)^{1/p}$.
%------------------------------------------------------------------------------------------------------------------
%------------------------------------------------------------------------------------------------------------------
\begin{figure}[tb]
\centering
\includegraphics[scale=0.5]{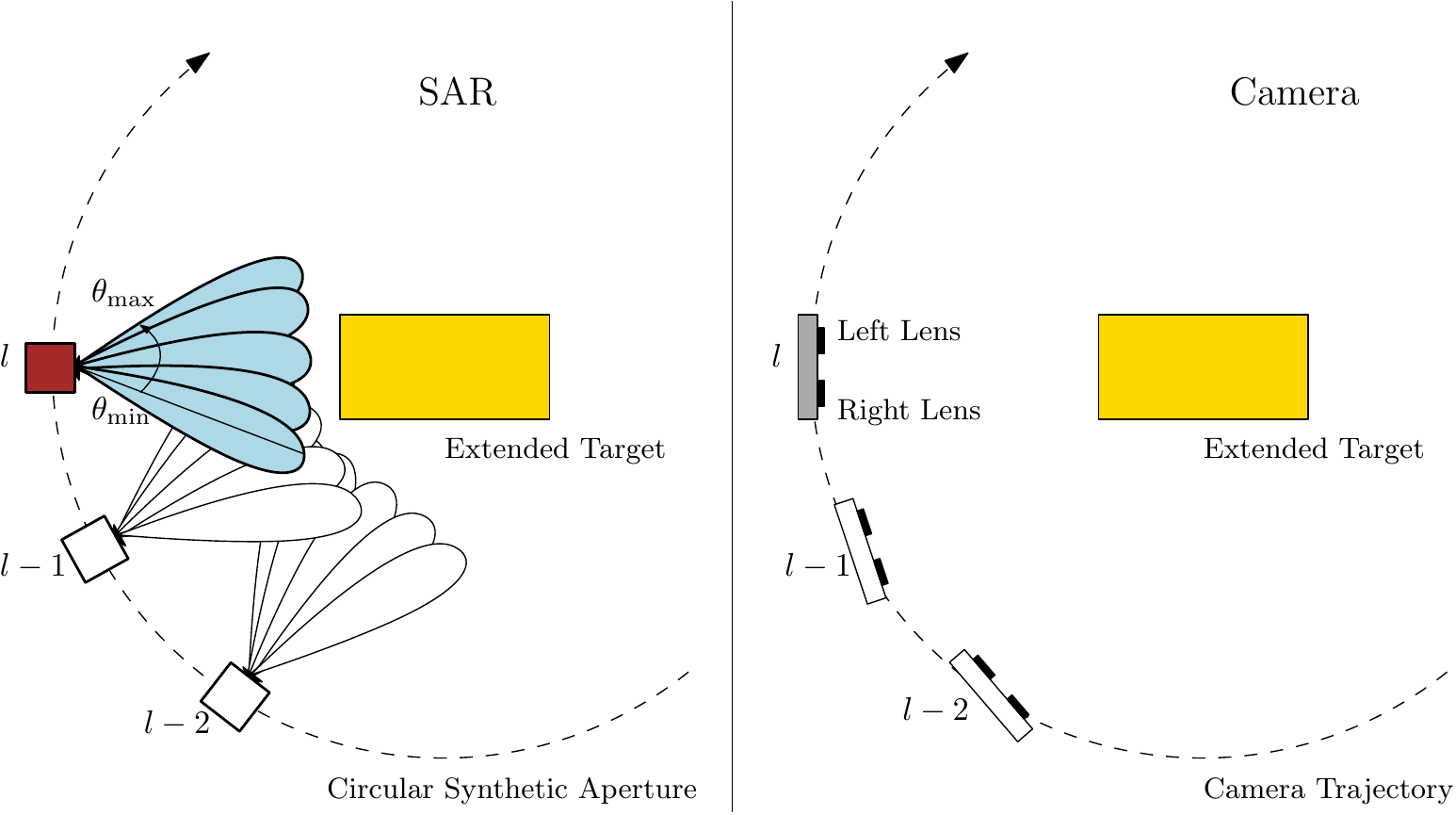} 
\caption{SAR and Camera System Schematic}
\label{fig:schem}
\end{figure}
%------------------------------------------------------------------------------------------------------------------
%------------------------------------------------------------------------------------------------------------------
\section{System Model}
\label{sect:sig_mod}
In \cite{SG_FSSAR_2018}, we proposed an FS-SAR mode to enhance the azimuth resolution of an automotive radar. This mode combines forward-scanning with SAR processing. FS-SAR assumes a linear aperture. In the present paper, we consider a circular aperture, i.e., a circular-scanning SAR (CiS-SAR). CiS-SAR combines the benefits of scanning and spotlight SAR, resulting in enhanced azimuth resolution. We opt for CiS-SAR as a generic SAR mode in order to exhibit the cross-learning possibilities from camera to SAR. However, future works may consider other SAR modes as well.
Figure~\ref{fig:schem} shows the schematic for CiS-SAR. At each scan-step, $l\in [1,L]$, over the circular aperture, the radar scans the extended target over the angular range (field of view of the sensor), $\theta \in [\theta_{\rm min}, \theta_{\rm max}]$. 
Thus, the target information is obtained both over the circular aperture as well as over the angular scan per aperture position. Similar to compressed sensing-based back-projection (CBP) in \cite{SG_FSSAR_2018}, we first process the measurements received over the scans by using a compressed sensing-based algorithm to improve the resolution and then back-project the reconstructed images from all the scans over the circular aperture to generate a coherent image of the extended target.

Let, the radar transmits frequency modulated continuous wave (FMCW) pulses towards the target. The signal received is then deramped, low-pass filtered, deskewed, and Fourier transformed along the fast-time to obtain the range profile (see \cite{SG_FSSAR_2018} for explicit expressions and subsequent details on the radar signal model). 
However, the received signal along the azimuth, for a scanning radar, at scan-step $l$ and range $r$ can be considered as a convolution of the radar antenna beam, $h(\theta)$, and the azimuth reflectivity function, $x_{l,r}(\theta)$, i.e., 
\be
{y}_{l,r}(\theta) = h(\theta) \star {x}_{l,r}(\theta) + \nu_{l,r}(\theta)
\label{eq:y_conv}
\ee
where $\star$ denotes convolution and $\nu_{l,r}(\theta)$ represents the model/thermal noise. Collecting all of the azimuth samples over $\theta$, (\ref{eq:y_conv}) can be written as
\be
\ybf_{l,r} = \Gbf {\Hbf} {\xbf}_{l,r} + \nubf_{l,r}
\label{eq:measur1}
\ee
where $\Hbf$ is a block-Toeplitz convolution matrix and $\Gbf$ is a selection matrix to balance the numeric relation between $N_x\times 1$ vector ${\xbf}_{l,r}$ and $N_\theta\times 1$ vector $\ybf_{l,r}$.
Note, in the context of azimuth-resolution enhancement, $N_x\gg N_\theta$.
Now, concatenating $\ybf_{l,r}$ over all $N_r$ range bins, we can write (\ref{eq:measur1}) as
\be
\ybf_l =  \underbrace{[\Ibf_{N_r} \otimes (\Gbf\Hbf)]}_{\;\;\; \defeq \; \Abf } \xbf_l + \nubf_l 
\label{eq:measurVec}
\ee
where $\ybf_l \defeq [\ybf_{l,1}^T, \ybf_{l,2}^T, \cdots, \ybf_{l,{N_r}}^T]^T$ is an $N_\theta N_r\times 1$ vector. Similarly, $\xbf_l$ and $\nubf_l$ can be defined as $N_xN_r\times 1$ and $N_\theta N_r\times 1$ vectors, respectively. 
Further, $\Abf$ is the $N_\theta N_r\times N_xN_r$ measurement matrix.
Now, according to CBP, $\xbf_l$ can be estimated by solving the following (fused LASSO \cite{fusedLasso}) optimisation problem.
\be
\hat{\xbf}_l  = \mathop{\arg \min}\limits_{\xbf_l}  \left\| \ybf_l - \Abf \xbf_l \right\|_2^2 + \lambda_e \left\| \xbf_l \right\|_1^1
+ \lambda_f \left\| \Dbf \xbf_l \right\|_1^1
\label{eq:fl}
\ee
where $\lambda_e$ and $\lambda_f$ are positive penalty parameters controlling element-wise sparsity and fusion in $\xbf_l$, respectively, and $\Dbf$ is the fusion matrix (i.e., $\Dbf \xbf_l$ is a vector of differences of consecutive elements of $\xbf_l$) \cite{fusedLasso}.
Then, the reconstructed radar image via back-projection, at pixel $(i,j)$, can mathematically be represented as
\be
\gamma_{i,j} =  \sum_{l=0}^{L-1} [ \uparrow_{1,\kappa'}(\hat{\Xbf}_l) ]_{I_{\theta_{i,j}},I_{r_{i,j}} }
\label{eq:gamma_ij_x}
\ee 
where
$\uparrow_{\kappa,\kappa'}(\cdot)$ interpolates/upsamples a matrix by an order $\kappa$ and $\kappa'$ along its rows and columns, respectively,
$\hat{\Xbf}_l$ is the $N_x\times N_r$ reshaped matrix form of $\hat{\xbf}_l$,
$[\cdot]_{I_{\theta_{i,j}},I_{r_{i,j}} }$ represents the row and column indices of the matrix corresponding to angle $\theta_{i,j}$ and range $r_{i,j}$ for the $(i,j)$th pixel, respectively. 
Since each scanning position over the aperture may contribute to each pixel in the reconstructed image, we can rewrite (\ref{eq:gamma_ij_x}) as 
\be
\gamma_{i,j} =  \sum_{l=0}^{L-1} \gamma_{i,j}^l
\label{eq:gamma_ij_x_}
\ee 
where $\gamma_{i,j}^l \defeq [ \uparrow_{1,\kappa'}(\hat{\Xbf}_l) ]_{I_{\theta_{i,j}},I_{r_{i,j}} }$.
Let, all of the image pixels $\gamma_{i,j}^l$, for $i=1,\cdots,\sqrt{N}$ and $j=1,\cdots,\sqrt{N}$, w.r.t. contributions from the $l$th aperture position, are collected in an $N\times 1$ vector $\rbf_l$, i.e., 
\be
\rbf_l \defeq \left[\gamma_{1,1}^l, \cdots, \gamma_{\sqrt{N},1}^l, \gamma_{\sqrt{N},2}^l, \cdots, \gamma_{\sqrt{N},\sqrt{N}}^l \right]^T.
\label{eq:r_l_img}
\ee
Now, we can collect all of the radar images generated from each aperture position, as defined in (\ref{eq:r_l_img}), as an $N\times L$ matrix $\Rbf$, i.e., 
\be
\Rbf \defeq \left[ \rbf_1, \rbf_2, \cdots, \rbf_L \right].
\label{eq:R_img}
\ee

For the proposed cross-learning, we assume that the camera trajectory is the same as that of the radar synthetic aperture, i.e., the camera images of the target are also obtained from the same physical location as that of the radar.
However, the camera does not involve any scanning and takes one snapshot for each location over its trajectory.
Figure~\ref{fig:schem} shows the schematic of camera image acquisition. For the sake of clarity, the schematic for camera has been drawn separately from SAR. However, in practice, both sensors may share the physical location.
The camera can be mono or stereo, with different image formats, i.e, RGB, greyscale, depth-map etc. It may also have its own requirements w.r.t. configuration, calibration, disparity/point-cloud formulation, etc. We assume that these pre-requisites have already been met. However, we do not assume any strict registration between the camera and the radar, as it is very difficult and our approach essentially tries to circumvents its need.

Let, a $\sqrt{M}\times \sqrt{M}$ generic camera image of the target at $l$th position on its trajectory is represented as an $M\times 1$ vector $\sbf_l$ via lexicographic ordering (column ordered).
Then, we can collect all such images for the complete trajectory into an $M\times L$ matrix $\Sbf$ as
\be
\Sbf \defeq \left[ \sbf_1, \sbf_2, \cdots, \sbf_L \right].
\label{eq:S_img}
\ee
The rest of the paper essentially deals with (\ref{eq:R_img}) and (\ref{eq:S_img}) in terms of proposing a cross-learning strategy.
%------------------------------------------------------------------------------------------------------------------
%------------------------------------------------------------------------------------------------------------------
\section{Cross-Learning}
\label{sect:xlearn}
In order to improve SAR images via cross-learning from camera images, we adopt a multi-level abstraction approach. 
From the test images, we first learn the respective manifolds. Secondly, we use CCA to generate a coherent subspace between the two modalities. Thirdly, we use LLE for neighbourhood embedding w.r.t. test images of the radar and camera. Finally, the processed radar image is projected from the manifold domain back to the image domain.
We name this approach as, multi-level CCA-based (ML-CCA) cross-learning.
%------------------------------------------------------------------------------------------------------------------
%------------------------------------------------------------------------------------------------------------------
\subsection{Suitable Manifold}
\label{sect:manifold}
As explained earlier, after cross-learning, we need to reconstruct the SAR image from low-dimensional space of the manifold domain to the high-dimensional space of the image domain. Therefore, non-linear manifolds cannot be used. In terms of linear manifolds, we opt for the classical PCA based manifolds. 

PCA represents data by using the directions of maximum variance. Thus, it requires computing the principal eigenvectors of the data covariance matrix. Now, assuming that the datasets in (\ref{eq:R_img}) and (\ref{eq:S_img}) are centred (i.e., the corresponding sample means have been subtracted from them), the covariance matrices of radar and camera datasets can be defined as, $\Cbf_r\defeq (1/L)\Rbf\Rbf^T $ and $\Cbf_s\defeq (1/L)\Sbf\Sbf^T$, respectively. 
The eigenvalue decomposition (EVD) of the covariance matrices can then be carried out as
\be
{\rm EVD}(\Cbf_r) = \Ubf_r \Sigmabf_r \Vbf_r^T
\label{eq:evd_r}
\ee
\be
{\rm EVD}(\Cbf_s) = \Ubf_s \Sigmabf_s \Vbf_s^T
\label{eq:evd_s}
\ee
where matrices $\Ubf_r$ and $\Ubf_s$ contain the left eigenvectors, matrices $\Vbf_r$ and $\Vbf_s$ contain the right eigenvectors, and matrices $\Sigmabf_r$ and $\Sigmabf_s$ contain the corresponding eigenvalues along their diagonals, for radar and camera, respectively.
The low-dimensional data representation essentially corresponds to projecting the data on a few significant eigenvectors. 
Let, $n$ and $m$ represent the number of significant eigenvectors (or subsequent principal components) for SAR and camera manifolds, respectively. 
Then, $\bar{\Ubf}_r \defeq [\Ubf_r]_{:,1:n}$ and $\bar{\Ubf}_s \defeq [\Ubf_s]_{:,1:m}$ are the $N\times n$ and $M\times m$ corresponding PCA-based projection matrices. The PCA-based projection coefficients can be obtained as
\be
\Pbf_r = \bar{\Ubf}_r^T \Rbf = \left[\pbf_{r_1}, \pbf_{r_2}, \cdots, \pbf_{r_L} \right]
\label{eq:pca_coeff_r}
\ee 
\be
\Pbf_s = \bar{\Ubf}_s^T \Sbf = \left[\pbf_{s_1}, \pbf_{s_2}, \cdots, \pbf_{s_L} \right]
\label{eq:pca_coeff_s}
\ee 
where $\pbf_{r_l} \defeq \bar{\Ubf}_r^T \rbf_l$, $\pbf_{s_l} \defeq \bar{\Ubf}_s^T \sbf_l$, and, $\Pbf_r$ and $\Pbf_s$ are $n\times L$ and $m\times L$ PCA coefficient matrices of SAR and camera manifolds, respectively.
 %------------------------------------------------------------------------------------------------------------------
%------------------------------------------------------------------------------------------------------------------
\subsection{Coherent Subspace}
\label{sect:cca}
CCA finds a low-dimensional coherent subspace between two datasets. In this paper, we consider a one-dimensional CCA subspace. 
Thus, in our case, CCA provides one basis vector for each dataset such that the correlation between the corresponding projection coefficients is maximised. Note, the datasets, in our case, correspond to PCA-based manifold coefficients, i.e., (\ref{eq:pca_coeff_r}) and (\ref{eq:pca_coeff_s}). 
Mathematically, we can estimate the CCA-based subspace by solving the following optimisation problem, as in \cite{Hardoon_CCA_2004}.
\begin{subequations}
\begin{align}
\left[ \hat{\bbf}_r, \hat{\bbf}_s \right] =  \mathop{\arg \max}\limits_{\bbf_r, \bbf_s} \;\; &\bbf_r^T \Qbf_{rs} \bbf_s 
\label{eq:cca_def_1} \\
\text{s.t.} \;\; &\bbf_r^T \Qbf_{r} \bbf_r = 1,  \, \bbf_s^T \Qbf_{s} \bbf_s = 1 
\label{eq:cca_def_2}
\end{align}
\label{eq:cca_def}%
\end{subequations}
where $\bbf_r$ and $\bbf_s$ are $n\times 1$ and $m\times 1$ canonical basis vectors for the radar- and camera-manifold datasets, respectively, and, $\Qbf_{r} \defeq (1/L) \Pbf_r \Pbf_r^T$, $\Qbf_{s} \defeq (1/L) \Pbf_s\Pbf_s^T$ and $\Qbf_{rs} \defeq (1/L) \Pbf_r \Pbf_s^T$ are the corresponding covariance matrices. Note, the constraints (\ref{eq:cca_def_2}) are imposed to ensure a unique solution. 
Now, solving (\ref{eq:cca_def}) essentially boils down to solving the following generalised eigenvalue problem (see \cite{Hardoon_CCA_2004} for details). 
\be
\begin{bmatrix}
\Qbf_{rs}^T & \zerobf \\
\zerobf & \Qbf_{rs}
\end{bmatrix}
\begin{bmatrix}
\bbf_r \\
\bbf_s
\end{bmatrix}
=
2\lambda
\begin{bmatrix}
\zerobf & \Qbf_{s} \\
\Qbf_{r} & \zerobf
\end{bmatrix}
\begin{bmatrix}
\bbf_r \\
\bbf_s
\end{bmatrix}
\label{eq:gen_eigen}
\ee
where $\lambda$ is the generalised eigenvalue.
Solving (\ref{eq:cca_def}) is equivalent to finding the largest generalised eigenvalue in (\ref{eq:gen_eigen}), i.e., $\lambda = \lambda_{\max}$, and the corresponding generalised eigenvector provides the estimate of canonical basis vectors as, $[\hat{\bbf}_r^T, \hat{\bbf}_s^T]^T$.
From the basis vectors, the corresponding CCA-based coefficients can be obtained as
\be
\abf_r =  \Pbf_r^T \hat{\bbf}_r
\label{eq:a_r}
\ee
\be
\abf_s =  \Pbf_s^T \hat{\bbf}_s
\label{eq:a_s}
\ee
where $\abf_r$ and $\abf_s$ are $L\times 1$ vectors of CCA-based coefficients w.r.t. the radar and camera manifolds, respectively.
%------------------------------------------------------------------------------------------------------------------
%------------------------------------------------------------------------------------------------------------------
\subsection{Neighbourhood Embedding}
\label{sect:lle}
LLE is used to compute low-dimensional neighbourhood-preserving embeddings of high-dimensional data.
It is based on a simple geometric intuition. Given a data point and its neighbours, in high-dimension, lie on a locally linear patch of the manifold, the data point can be reconstructed by linear combination of its neighbours. Then, the data point can be mapped to a low-dimensional representation while preserving its neighbourhood characterisation (see \cite{Roweis_00} for more details). 
In the context of cross-learning, {\it i)} the mapping is done from the camera manifold to the radar manifold, {\it ii)} the radar and the camera manifolds have been substituted with respective CCA-based coefficients which are linear due to one-dimensional CCA subspace and {\it iii)} in terms of CCA-based coefficients, the data dimension for both radar and camera is the same, therefore, we do not need to find the low-dimensional values.
Thus, LLE can be easily applied to our case for neighbourhood embedding, i.e., we need to find the linear coefficients which reconstruct a camera data point from its neighbours and then use the same linear coefficients to reconstruct a radar data point from its neighbours. This constitutes neighbourhood embedding in the context of cross-learning.

Let, $\rbf_t$ and $\sbf_t$ be the test SAR and camera images, respectively, with $\pbf_{r_t} \defeq \bar{\Ubf}_r^T \rbf_t$ and $\pbf_{s_t} \defeq \bar{\Ubf}_s^T \sbf_t$ as the corresponding data points on the manifolds. 
Then, the CCA-based coefficients for the test images can be obtained as
\be
a_{r_t} =  \pbf_{r_t}^T \hat{\bbf}_r
\label{eq:a_r_t}
\ee
\be
a_{s_t} =  \pbf_{s_t}^T \hat{\bbf}_s
\label{eq:a_s_t}
\ee
where $a_{r_t}$ and $a_{s_t}$ are scalar values.
Let, $\calN_{r_t}^K$ and $\calN_{s_t}^K$ represent the sets of $K$ nearest neighbours (K-NN) of $a_{r_t}$ and $a_{s_t}$, respectively. Now, we can write the optimisation problem of finding the linear coefficients of reconstructing $a_{s_t}$ from its neighbours in $\abf_s$ as a constrained least-squares fitting problem, i.e., 
\begin{subequations}
\begin{align}
\hat{\wbf} =  \mathop{\arg \min}\limits_{\wbf} \;\; & \| a_{s_t} - \wbf^T \bar{\abf}_s \|_2^2 
\label{eq:rec_s_t_1} \\
\text{s.t.} \;\; & \| \wbf \|_2^2 = 1
\label{eq:rec_s_t_2}
\end{align}
\label{eq:rec_s_t}%
\end{subequations}
where $\bar{\abf}_s$ is $K\times 1$ sub-vector of $\abf_s$, such that, $[\abf_s]_i \in \calN_{s_t}^K$, for $i=1,\cdots,K$. Note, (\ref{eq:rec_s_t}) essentially applies two constraints, {\it i)} a sparseness constraint, i.e., weights are non-zero only for the K-NN of $a_{s_t}$, {\it ii)} an invariance constraint, i.e., the sum of linear coefficients equals one, as (\ref{eq:rec_s_t_2}). 
An efficient way to minimise the error in (\ref{eq:rec_s_t_1}) is to solve the following system of linear equations
\be
\Gbf \wbf = \unobf
\label{eq:linsys}
\ee
where $\Gbf \defeq (a_{s_t} - \bar{\abf}_s)(a_{s_t} - \bar{\abf}_s)^T$, and then rescale the coefficients to satisfy (\ref{eq:rec_s_t_2}) (more details in \cite{Roweis_00}).
Now, the learnt coefficients can be used to reconstruct the radar data point from the neighbours, i.e., 
\be
\hat{a}_{r_t} =  \hat{\wbf}^T \bar{\abf}_r
\label{eq:rec_r_t}
\ee
where $\bar{\abf}_r$ is $K\times 1$ sub-vector of $\abf_r$, such that, $[\abf_r]_i \in \calN_{r_t}^K$, for $i=1,\cdots,K$.
%------------------------------------------------------------------------------------------------------------------
%------------------------------------------------------------------------------------------------------------------
\subsection{Image Reconstruction}
\label{sect:recon}
After learning the CCA-based coefficient, the learnt radar image in the manifold domain can be obtained as
\be
\tilde{\pbf}_{r_t} = (\hat{\bbf}_r^T)^\dagger \hat{a}_{r_t}  + \pbf_{r_t} 
\label{eq:rec_r_pca}
\ee
where $(\cdot)^\dagger$ denotes the Moore-Penrose (or pseudo) inverse. Now, the radar image can be projected from the manifold domain back to the image domain as
\be
\tilde{\rbf}_t = \bar{\Ubf}_r \tilde{\pbf}_{r_t}
\label{eq:rec_r_img}
\ee
where $\tilde{\rbf}_t $ is the improved SAR image obtained via cross-learning from the camera image.
%------------------------------------------------------------------------------------------------------------------
%------------------------------------------------------------------------------------------------------------------
\begin{figure}[t]
\centering
  \begin{subfigure}[t]{0.49\linewidth}
    \centering
      \includegraphics[width=.99\textwidth]{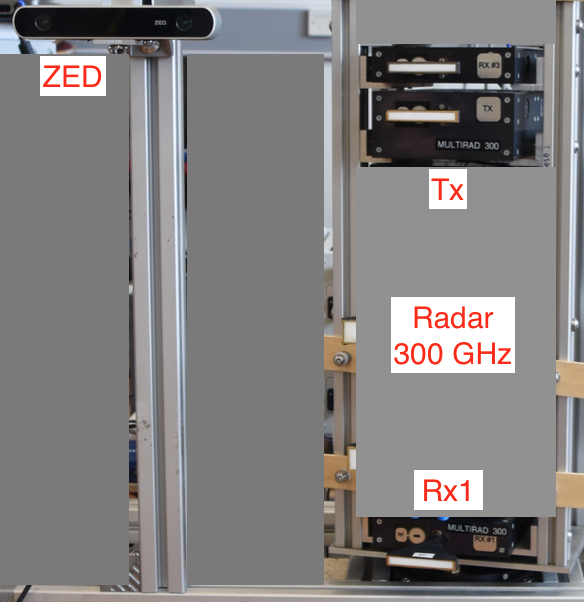} 
    \caption{ZED and Radar (300 GHz)}
    \label{fig:Rad_ZED}
  \end{subfigure}
  \begin{subfigure}[t]{0.49\linewidth}
    \centering
      \includegraphics[width=.99\textwidth]{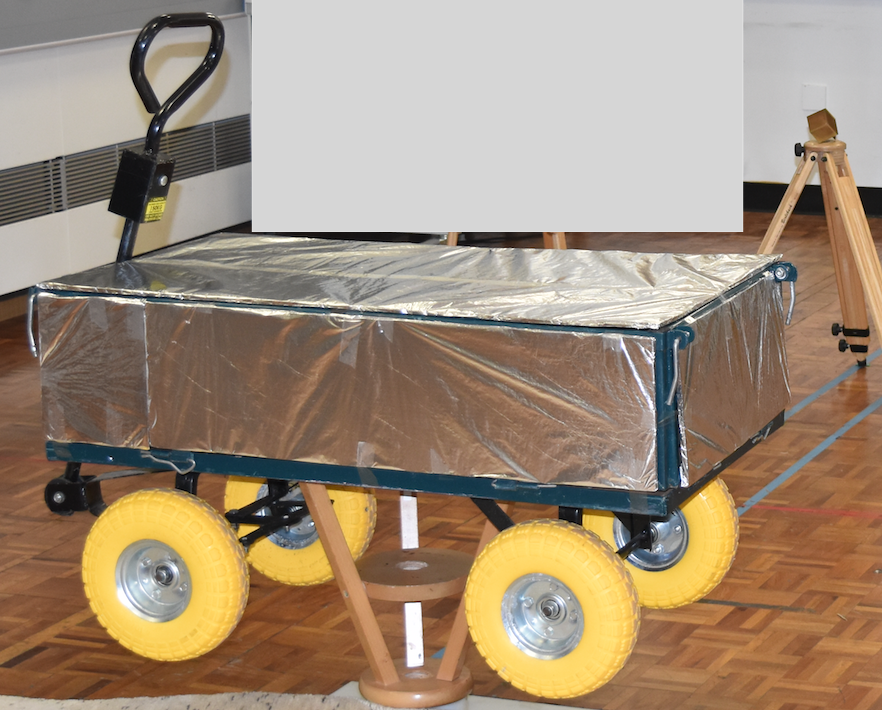} 
    \caption{Trolley on a Turn-Table}\label{fig:Trol_Tabl}
  \end{subfigure}
  \vspace{0.15in}
  \caption{Experimental Setup}
  \label{fig:Ex_setup}
\end{figure}
%------------------------------------------------------------------------------------------------------------------
\begin{table}
\renewcommand{\arraystretch}{1.3}
\begin{center}
\caption{Specifications of $300$ GHz Radar}
\begin{tabular}{ l | c}
\hline
Modulation & FMCW\\ \hline
Frequency Range & $287 - 293$ GHz\\ \hline
Transmit Bandwidh ($B$) & $5$ GHz\\ \hline
Chirp Duration ($T$) & $1$ ms\\ \hline
Sampling Frequency & $4.096$ MHz\\ \hline
Angular Step ($\Delta_\theta$) & $0.25^\circ$ \\ \hline
Range Resolution ($\Delta_r$) & $0.03$ m\\ \hline
Two-way $3$ dB Beamwidth ($\theta_{\rm 3dB}$) & $1.3^\circ$ \\ \hline
\end{tabular}
\label{tab:300}
\end{center}
\end{table}
%------------------------------------------------------------------------------------------------------------------
\begin{figure}[t]
\centering
      \includegraphics[scale=0.8]{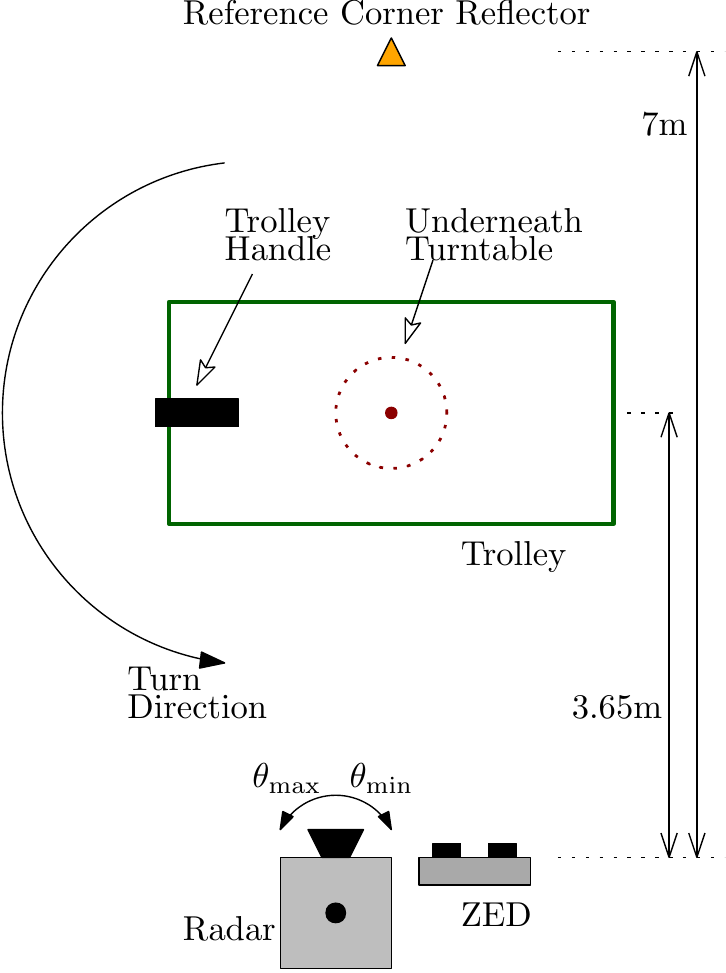} 
    \caption{Measurement Schematic}\label{fig:CiS_Msur_schem}
\end{figure}
%------------------------------------------------------------------------------------------------------------------
\begin{figure}[t]
\centering
  \begin{subfigure}[t]{0.49\linewidth}
    \centering
      \includegraphics[width=.99\textwidth]{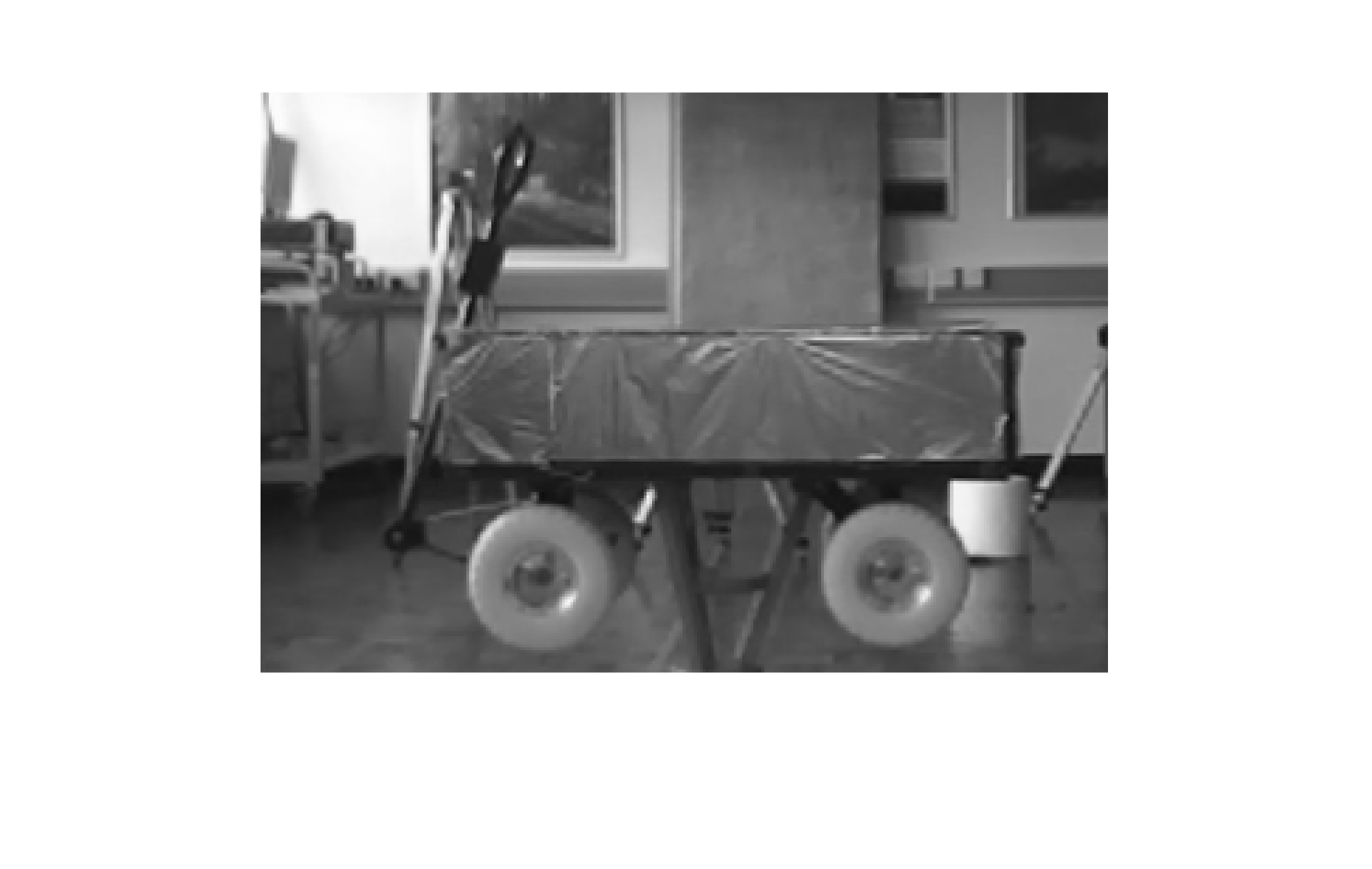} 
      \vspace{-0.35in}
    \caption{$l=1 \, (0^\circ)$}
    \label{fig:greyScale_img_1}
  \end{subfigure}
  \begin{subfigure}[t]{0.49\linewidth}
    \centering
      \includegraphics[width=.99\textwidth]{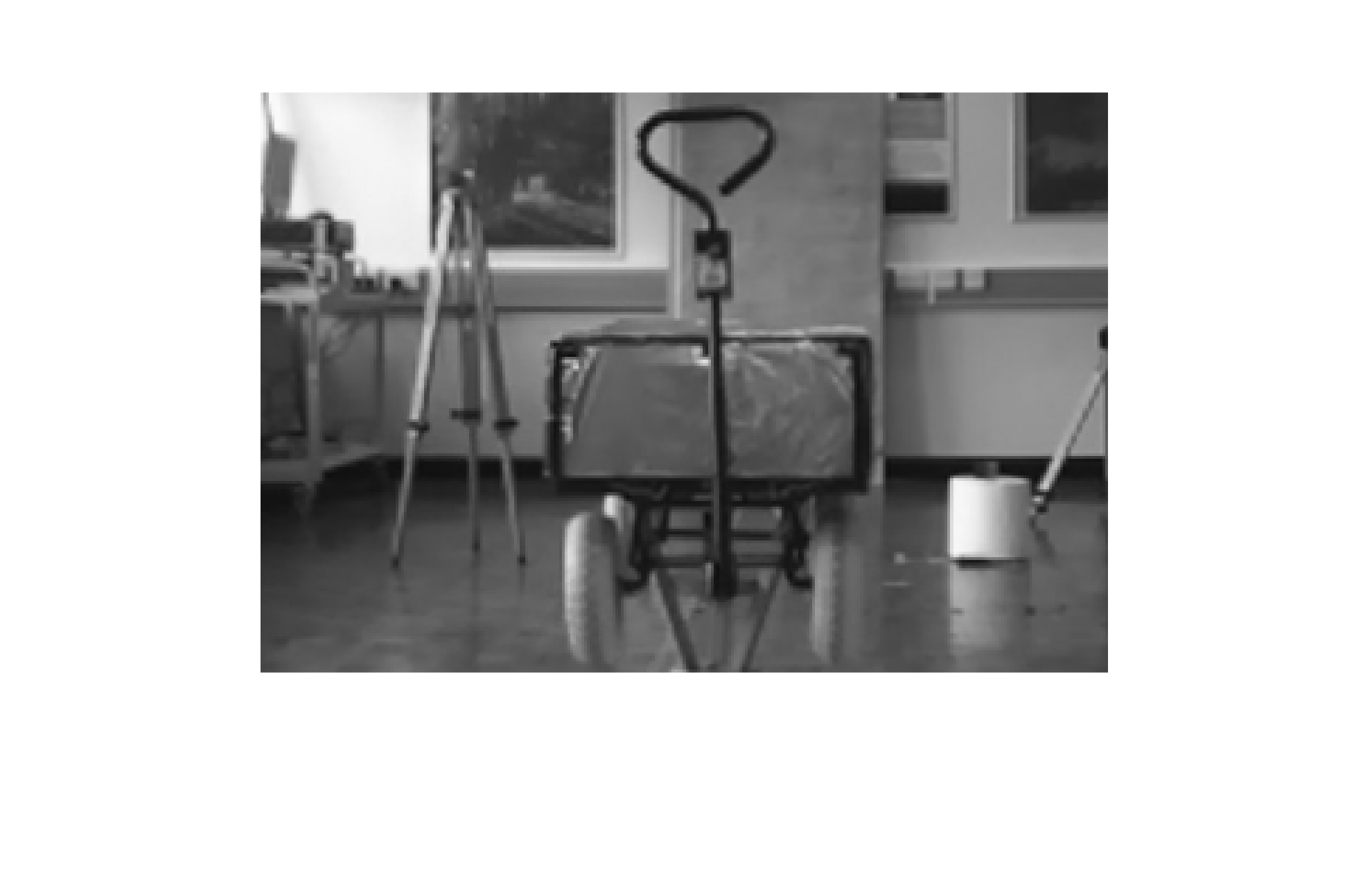} 
            \vspace{-0.35in}
    \caption{$l=19 \, (90^\circ)$}
    \label{fig:greyScale_img_19}
  \end{subfigure}
  \begin{subfigure}[t]{0.49\linewidth}
    \centering
      \includegraphics[width=.99\textwidth]{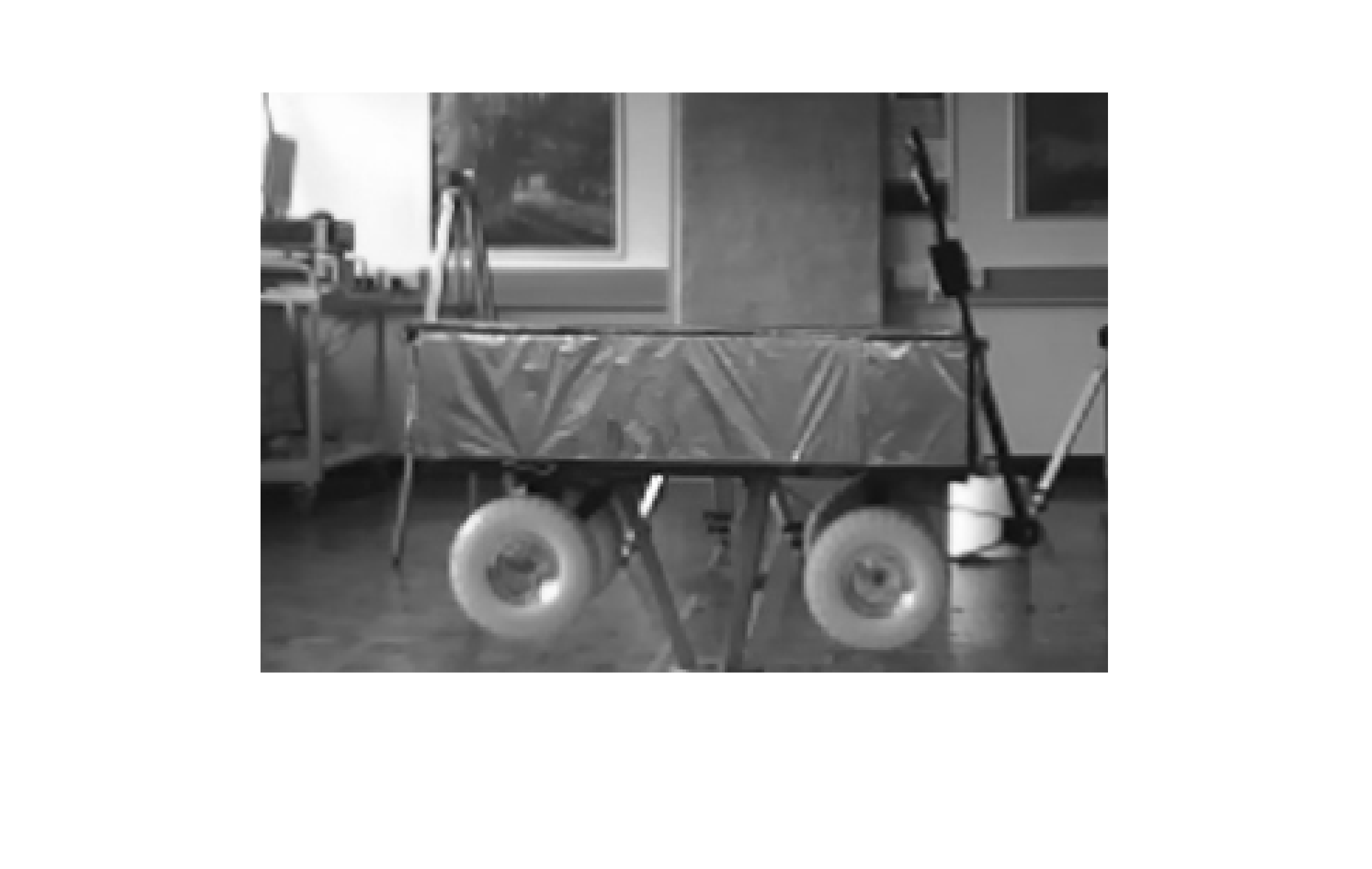} 
            \vspace{-0.35in}
    \caption{$l=37 \, (180^\circ)$}
    \label{fig:greyScale_img_37}
  \end{subfigure}
  \begin{subfigure}[t]{0.49\linewidth}
    \centering
      \includegraphics[width=.99\textwidth]{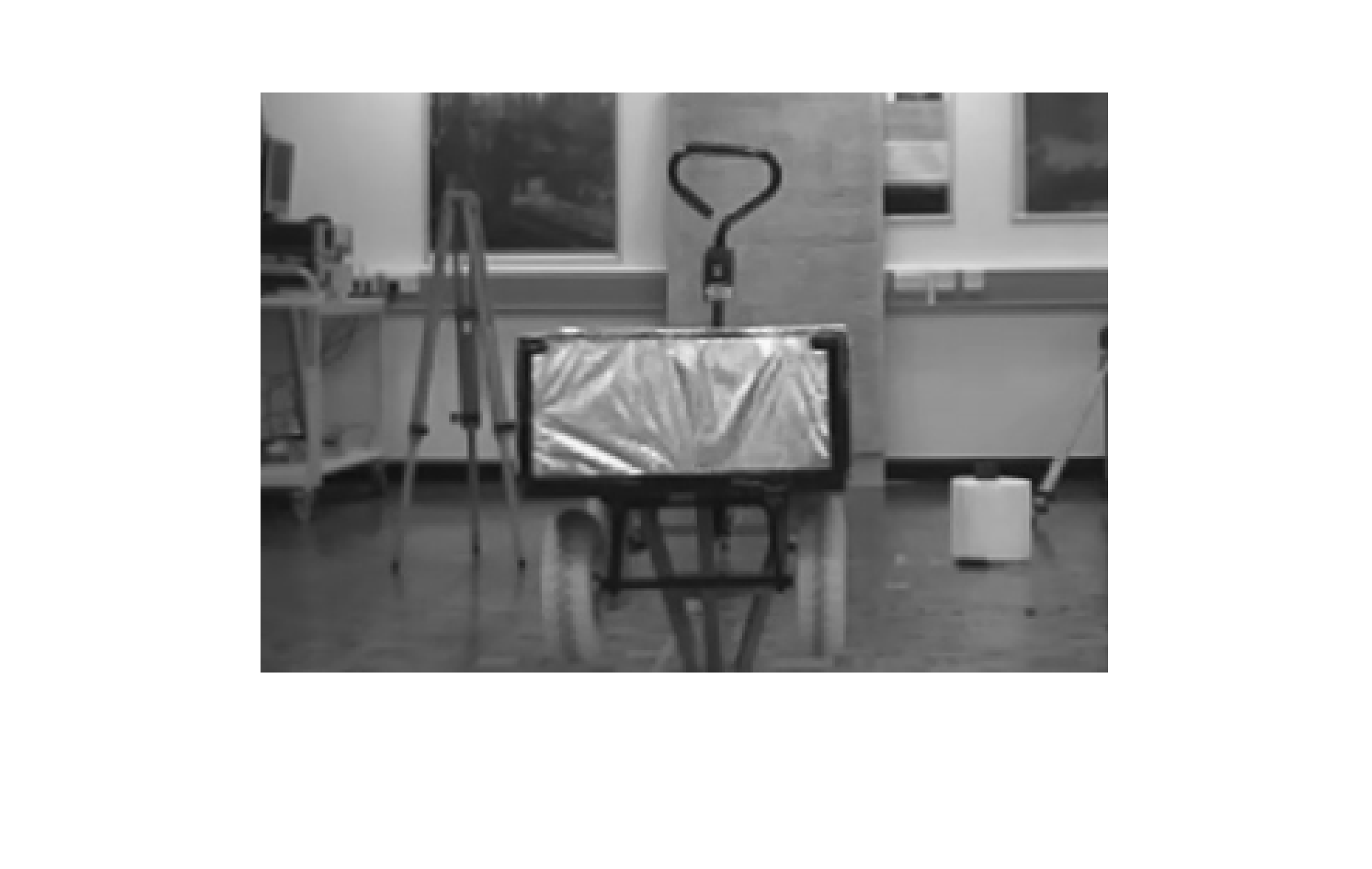} 
            \vspace{-0.35in}
    \caption{$l=55 \, (270^\circ)$}
    \label{fig:greyScale_img_55}
  \end{subfigure}
  \caption{ZED Greyscale Images}
  \label{fig:greyScale_img}
\end{figure}
%------------------------------------------------------------------------------------------------------------------
\begin{figure}[t]
\centering
  \begin{subfigure}[t]{0.49\linewidth}
    \centering
      \includegraphics[width=.99\textwidth]{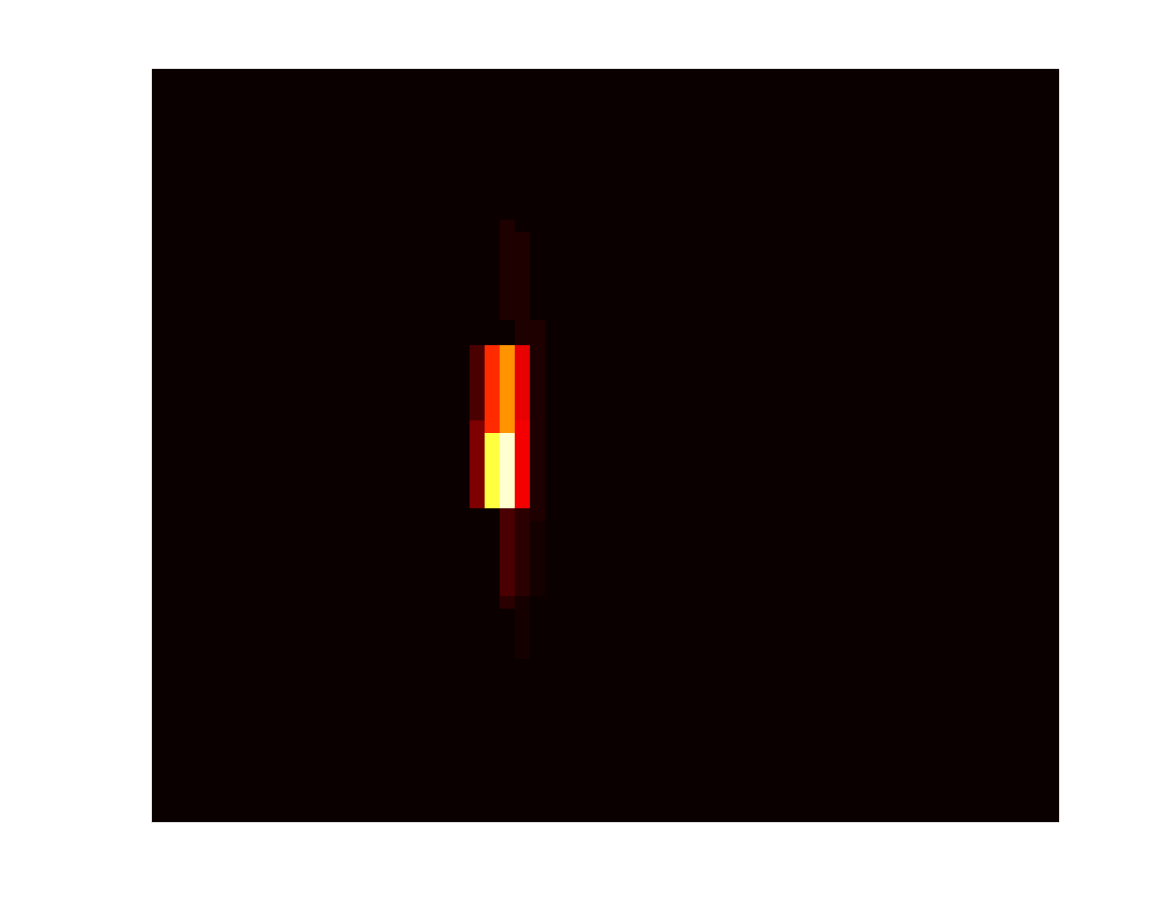} 
    \caption{$l=1\,(0^\circ)$}
    \label{fig:sar_1}
  \end{subfigure}
  \begin{subfigure}[t]{0.49\linewidth}
    \centering
      \includegraphics[width=.99\textwidth]{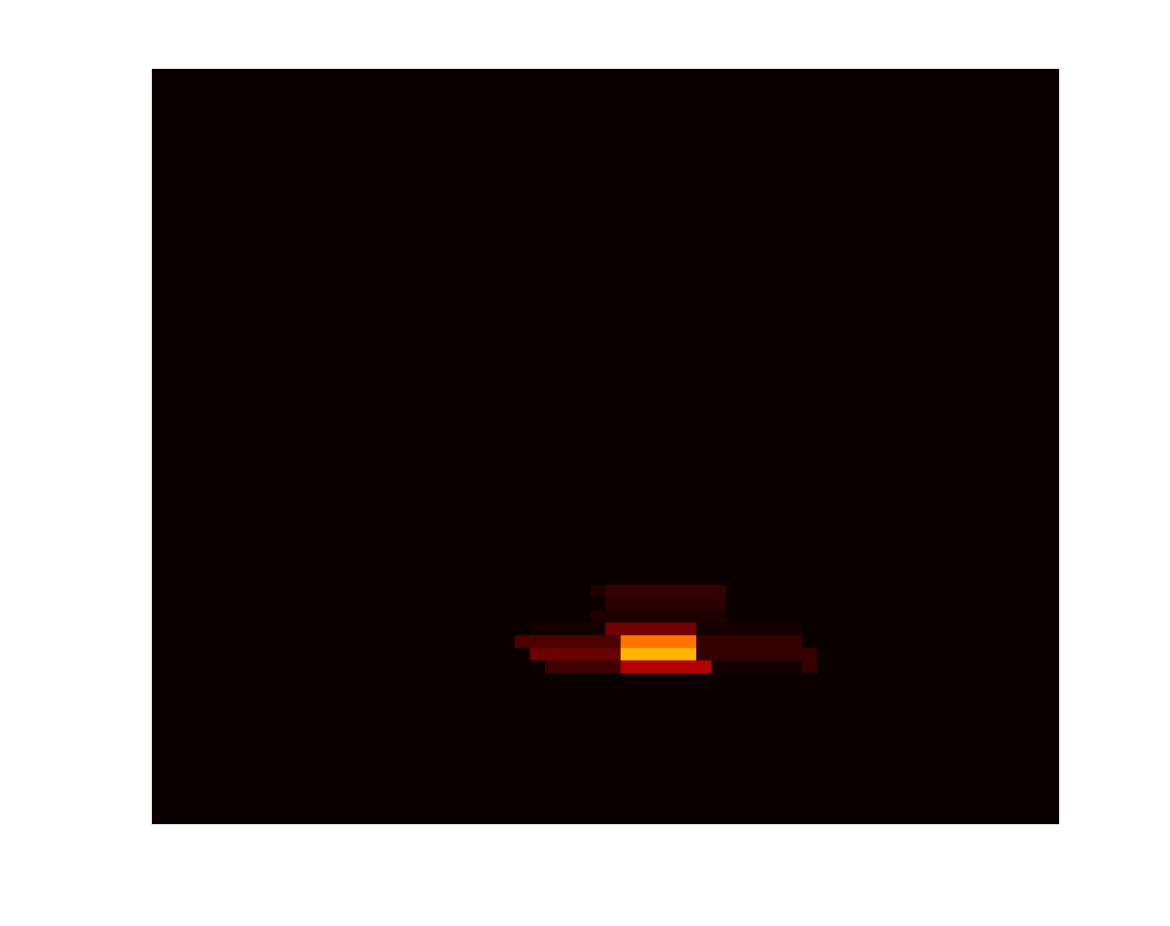} 
    \caption{$l=19\,(90^\circ)$}
    \label{fig:sar_19}
  \end{subfigure}
  \begin{subfigure}[t]{0.49\linewidth}
    \centering
      \includegraphics[width=.99\textwidth]{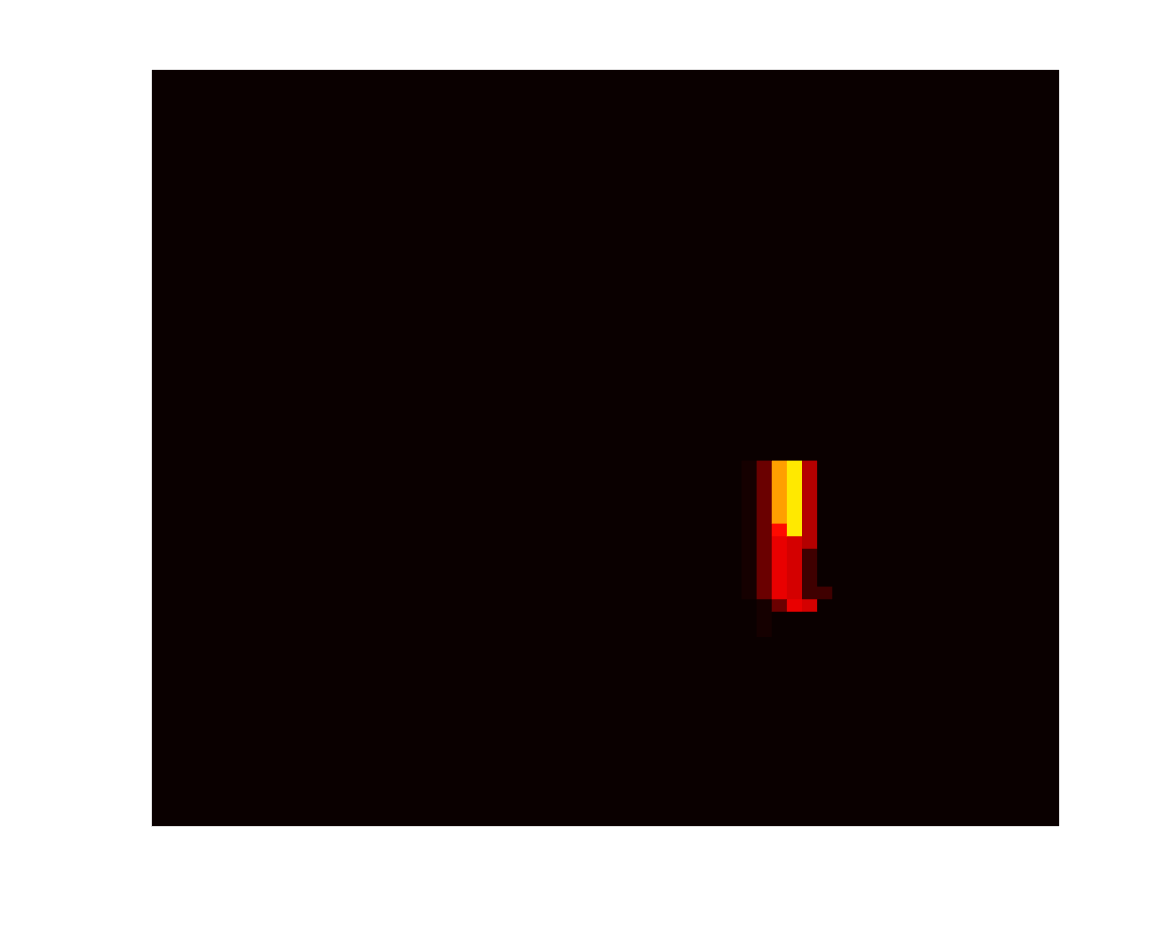} 
    \caption{$l=37\,(180^\circ)$}
    \label{fig:sar_37}
  \end{subfigure}
  \begin{subfigure}[t]{0.49\linewidth}
    \centering
      \includegraphics[width=.99\textwidth]{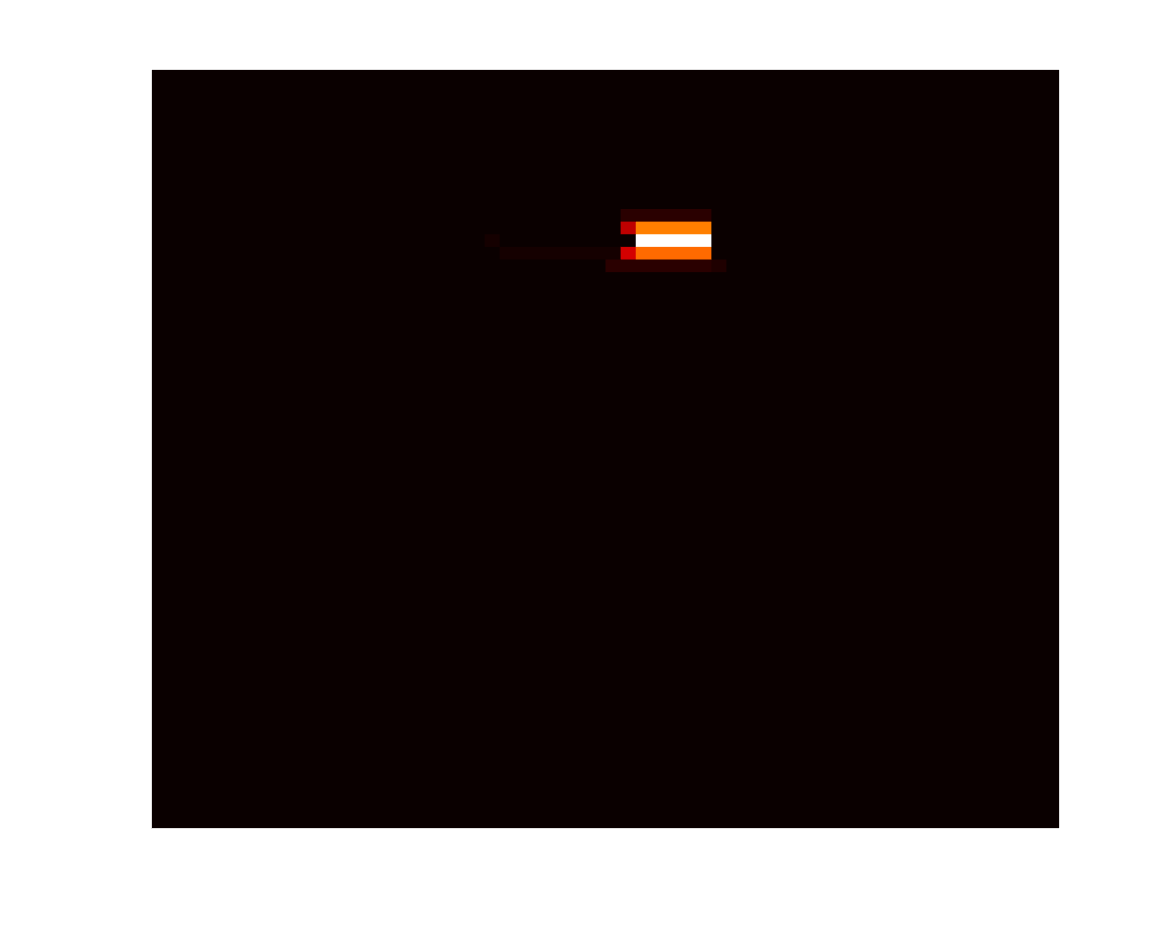} 
    \caption{$l=55\,(270^\circ)$}
    \label{fig:sar_55}
  \end{subfigure}
  \begin{subfigure}[t]{0.49\linewidth}
    \centering
      \includegraphics[width=.99\textwidth]{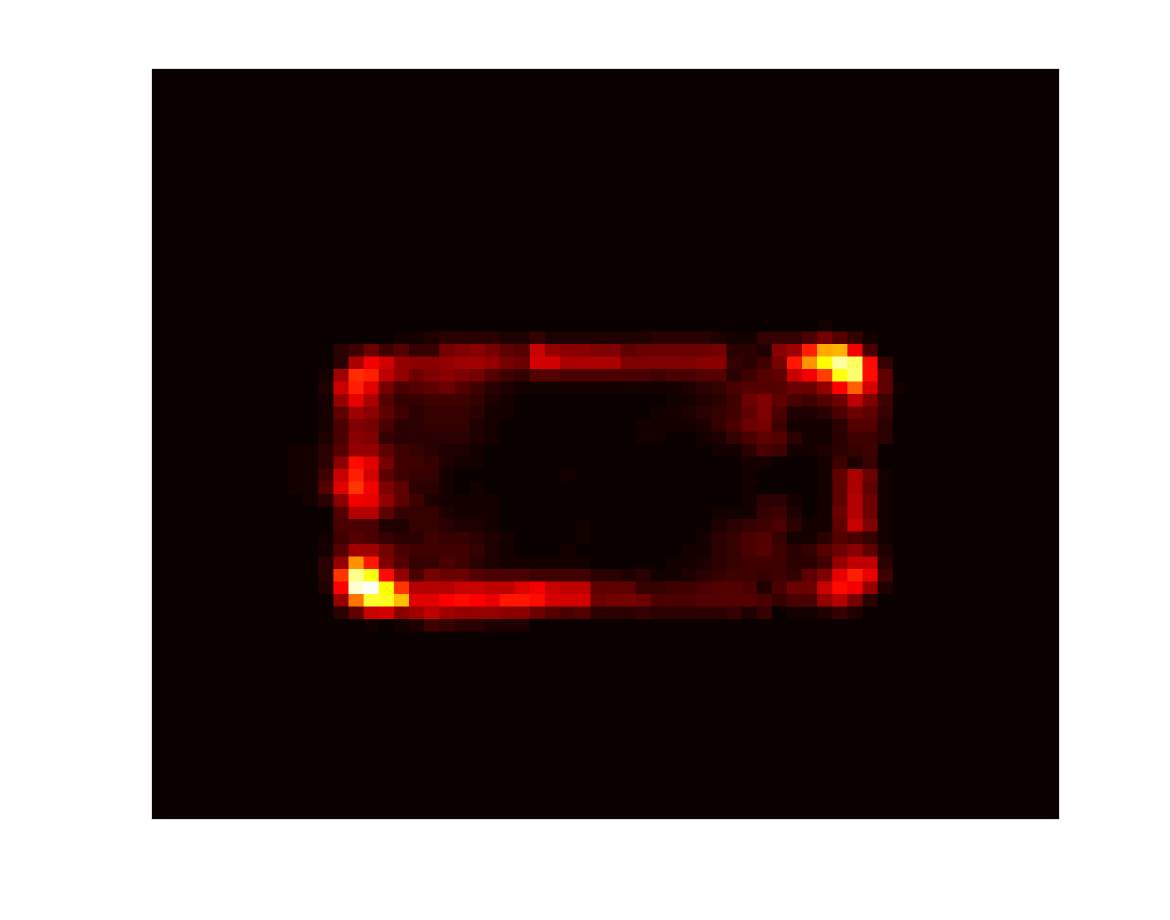} 
    \caption{CiS-SAR}
    \label{fig:mbc_original}
  \end{subfigure} 
  \caption{SAR Images}
  \label{fig:sar_img}
\end{figure}
%------------------------------------------------------------------------------------------------------------------
\begin{figure}[t]
\centering
  \begin{subfigure}[t]{0.49\linewidth}
    \centering
      \includegraphics[width=.99\textwidth]{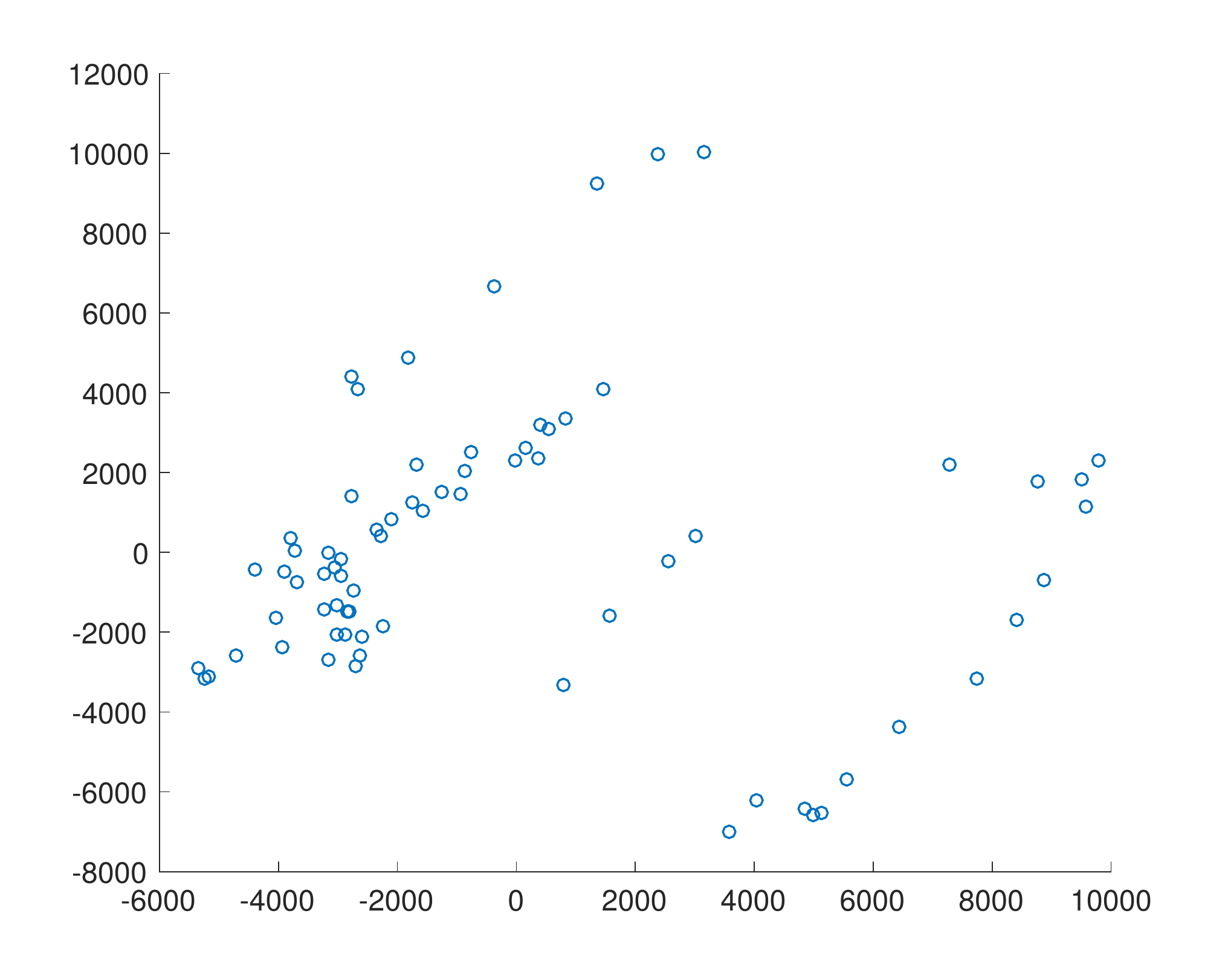} 
    \caption{Camera Manifold}
    \label{fig:greyScale_pca}
  \end{subfigure}
  \begin{subfigure}[t]{0.49\linewidth}
    \centering
      \includegraphics[width=.99\textwidth]{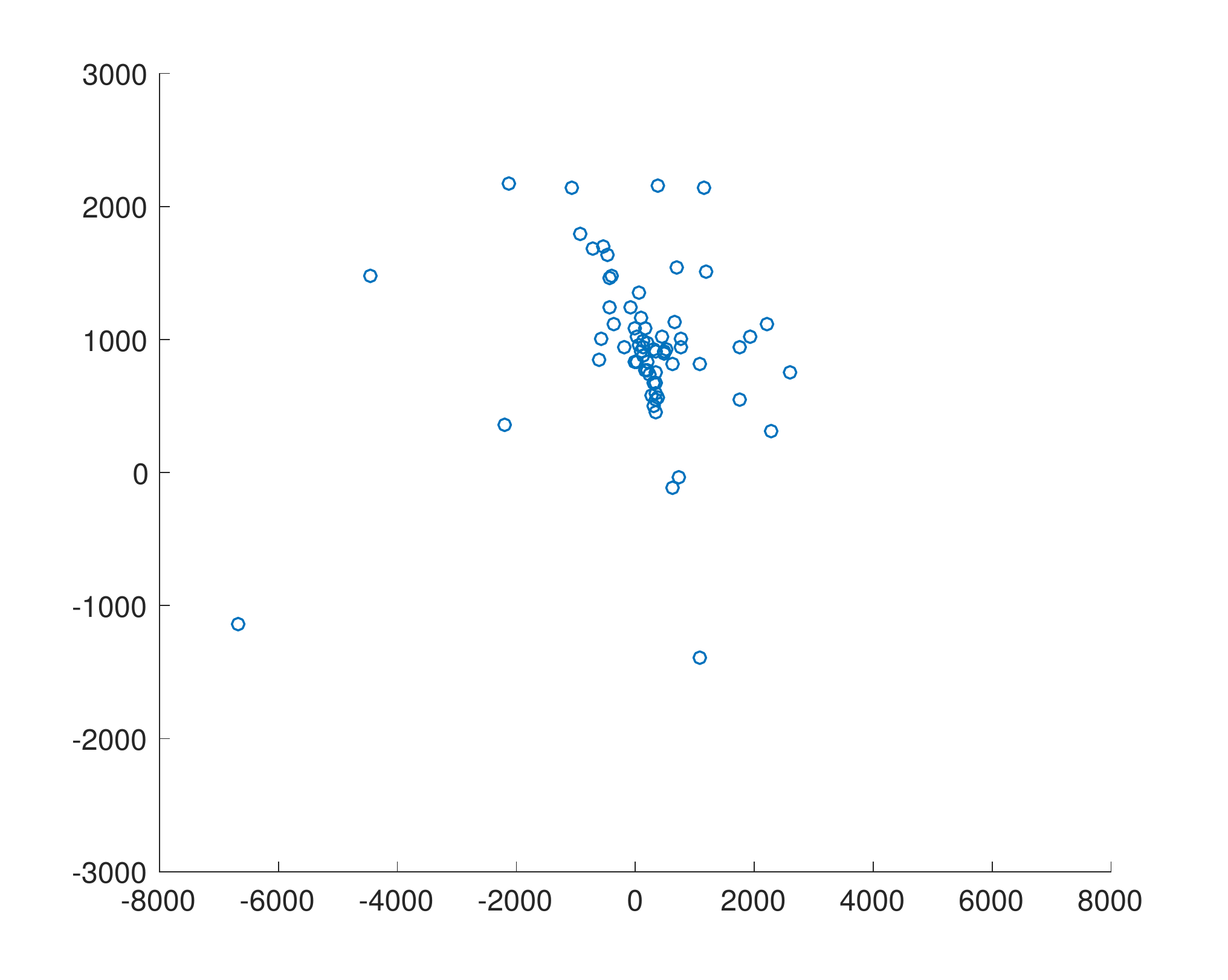} 
    \caption{SAR Manifold}
    \label{fig:mbc_pca}
  \end{subfigure}
  \caption{PCA-based Manifolds}
  \label{fig:pca}
\end{figure}
%------------------------------------------------------------------------------------------------------------------
\begin{figure}[t]
\centering
  \begin{subfigure}[t]{0.49\linewidth}
    \centering
      \includegraphics[width=.99\textwidth]{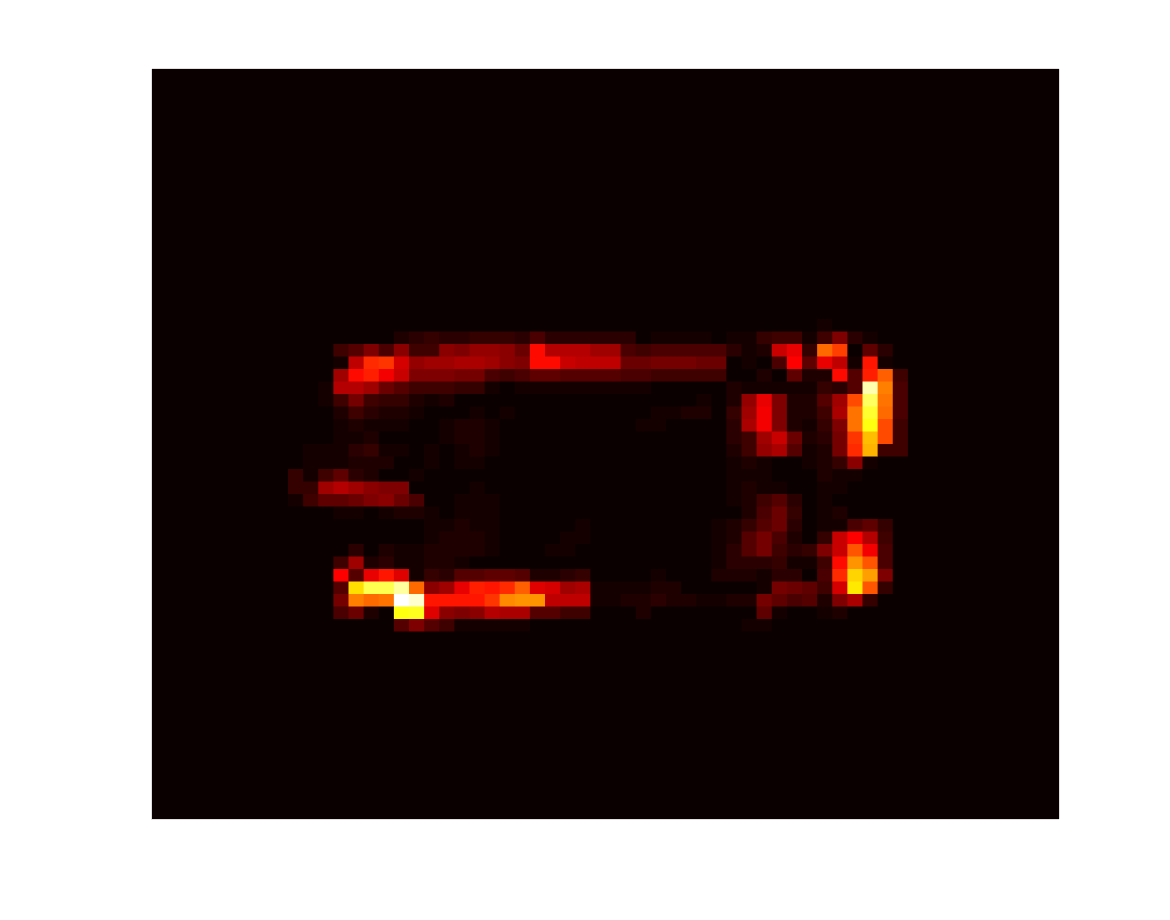} 
    \caption{ML-CCA}
    \label{fig:mbc_cca}
  \end{subfigure}
  \begin{subfigure}[t]{0.49\linewidth}
    \centering
      \includegraphics[width=.99\textwidth]{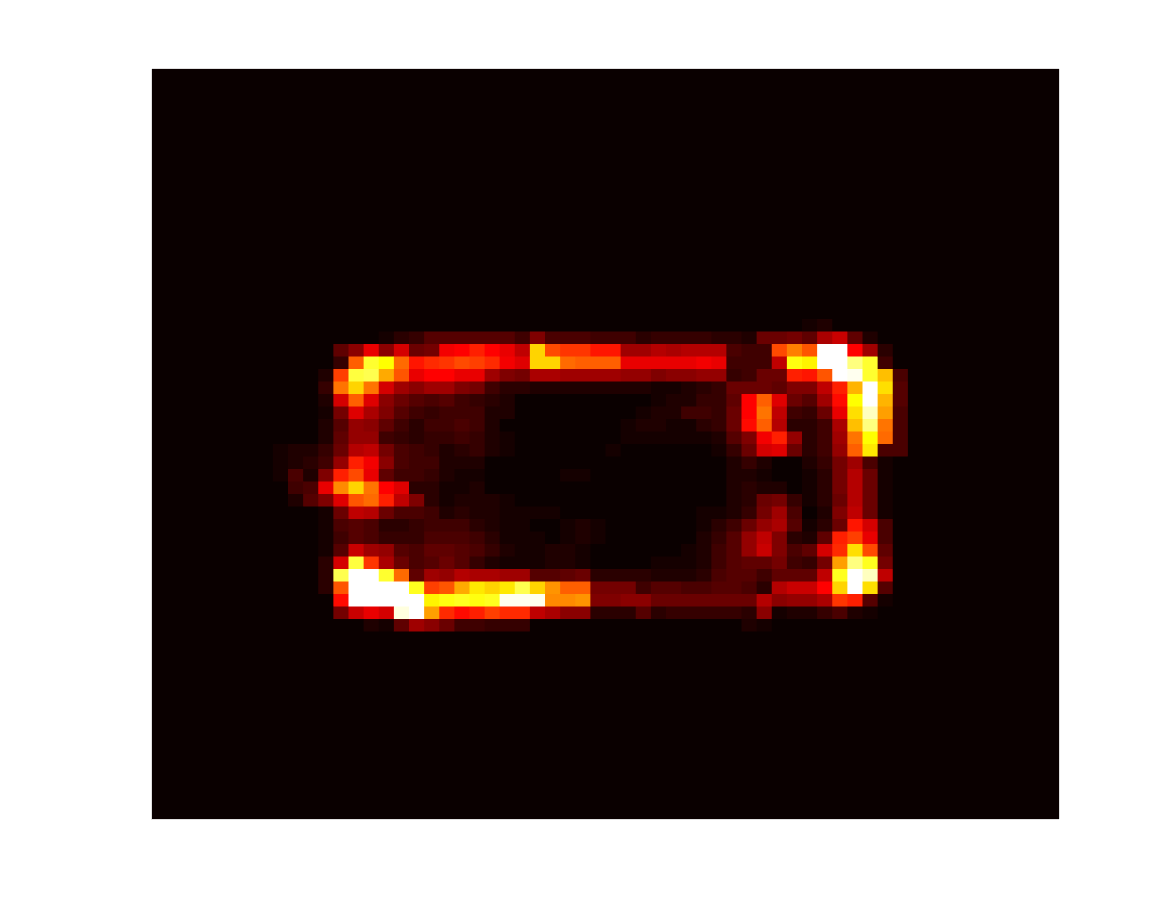} 
    \caption{ML-CCA+}
    \label{fig:mbc_cca_orig}
  \end{subfigure}
  \begin{subfigure}[t]{0.49\linewidth}
    \centering
      \includegraphics[width=.99\textwidth]{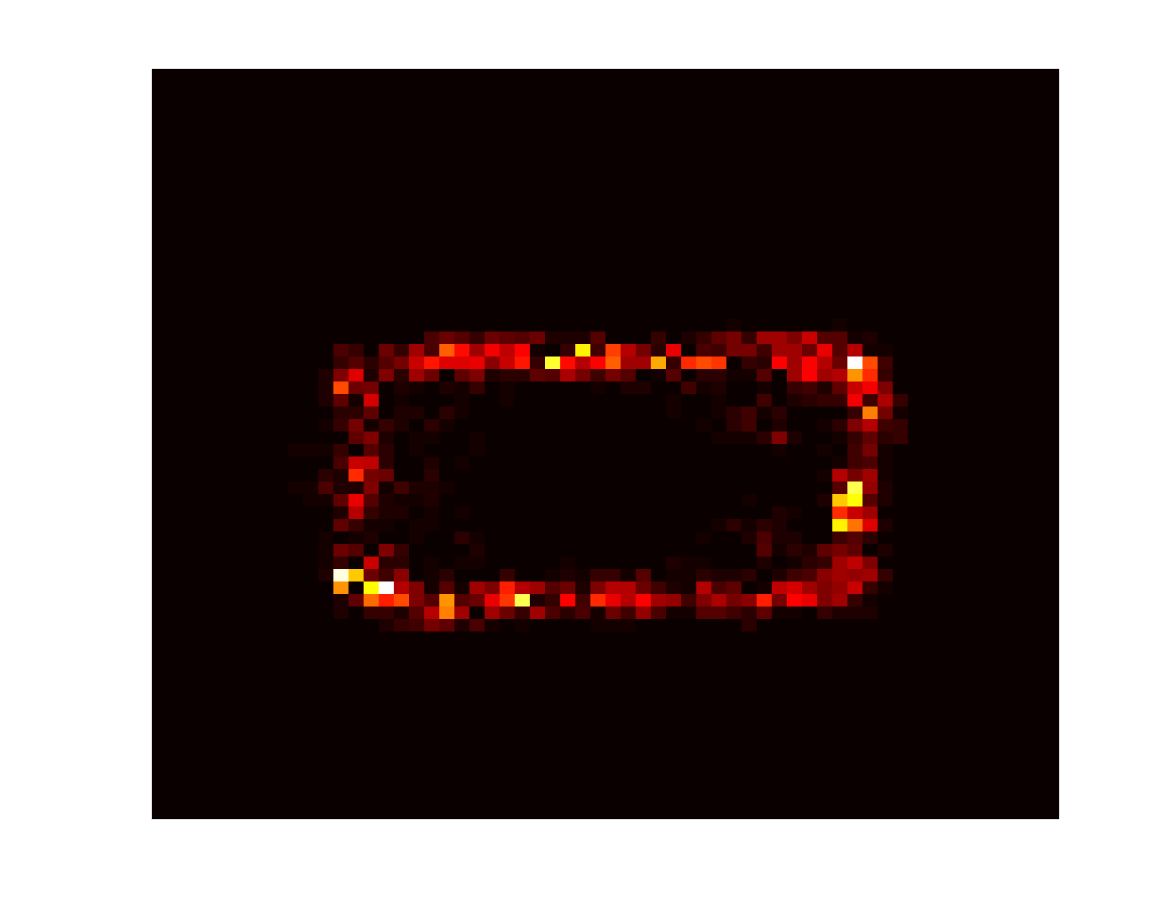} 
    \caption{MFA}
    \label{fig:mbc_maniAlign}
  \end{subfigure}
  \begin{subfigure}[t]{0.49\linewidth}
    \centering
      \includegraphics[width=.99\textwidth]{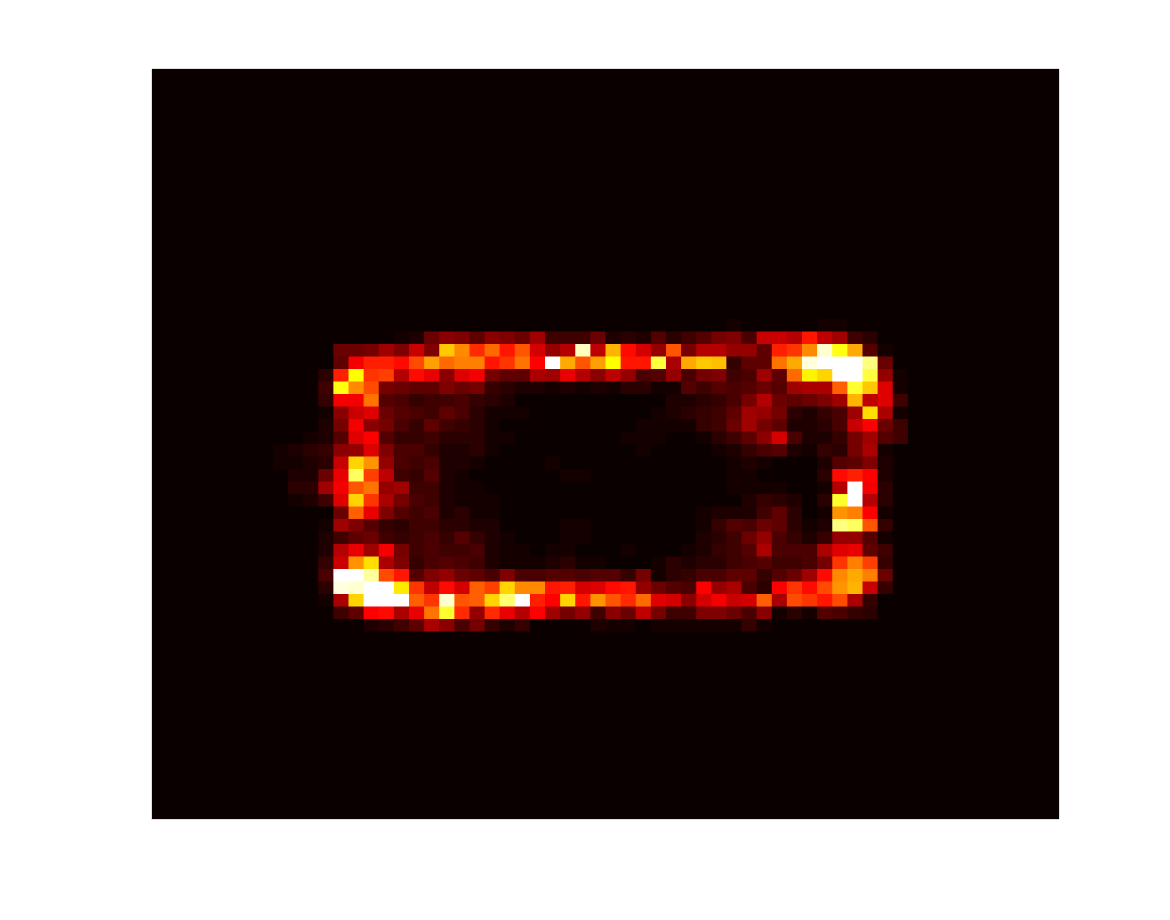} 
    \caption{MFA+}
    \label{fig:mbc_maniAlign_orig}
  \end{subfigure}
  \begin{subfigure}[t]{0.49\linewidth}
    \centering
      \includegraphics[width=.99\textwidth]{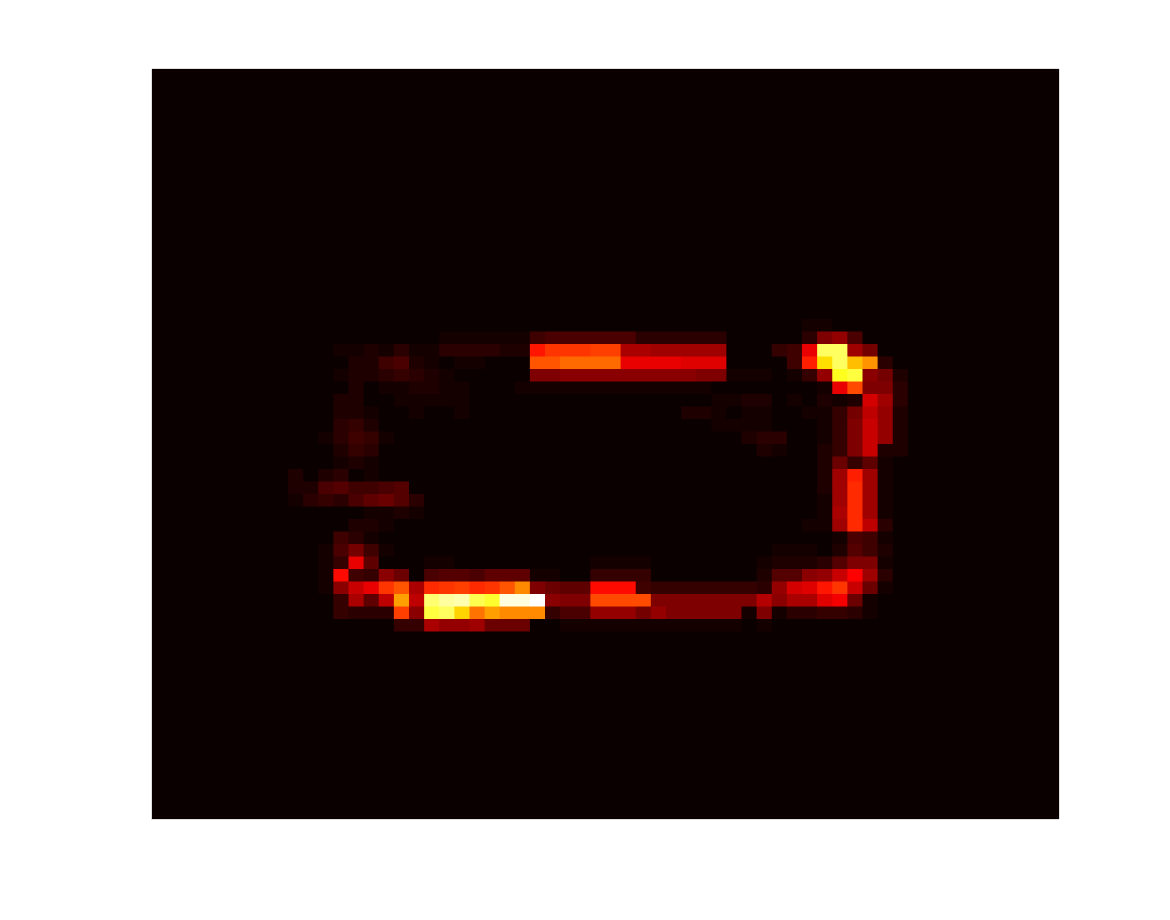} 
    \caption{ML-MFA}
    \label{fig:mbc_maniAlignLLE}
  \end{subfigure}
  \begin{subfigure}[t]{0.49\linewidth}
    \centering
      \includegraphics[width=.99\textwidth]{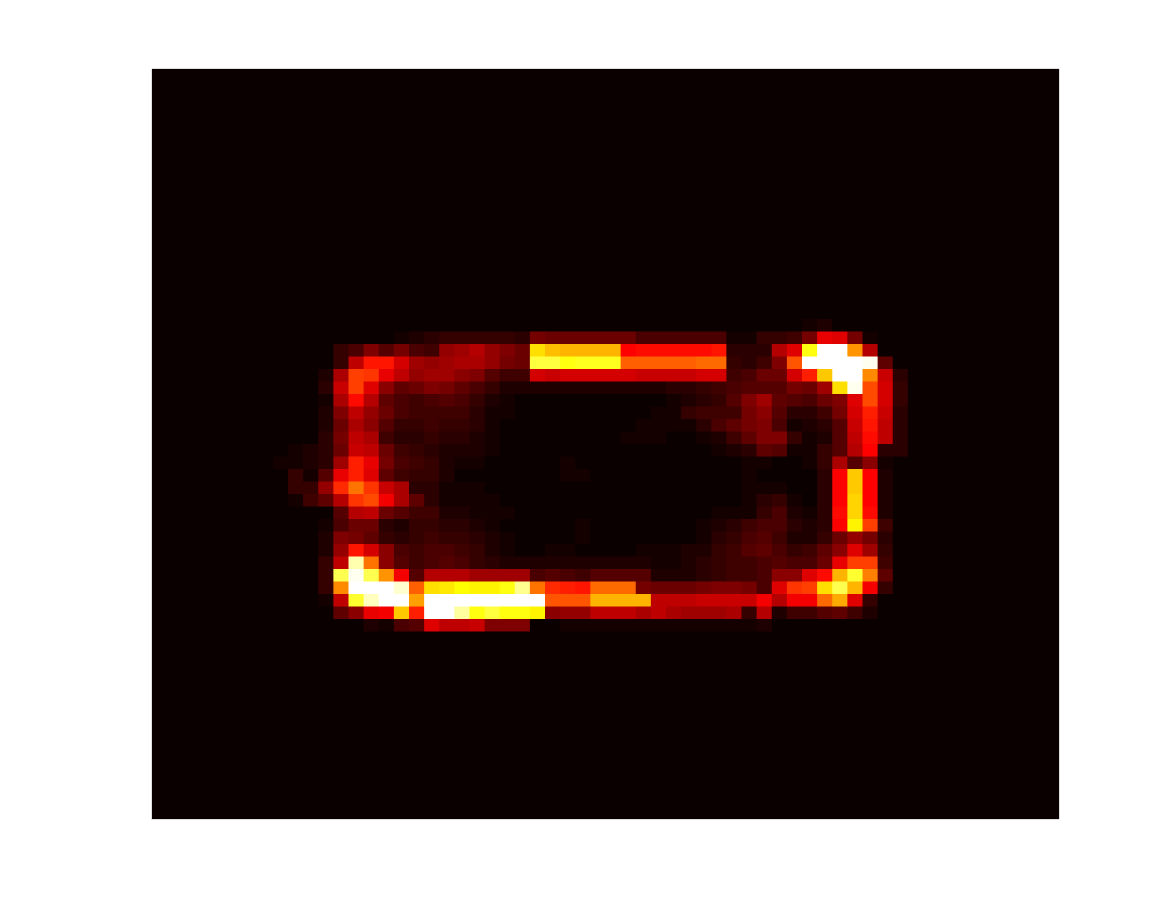} 
    \caption{ML-MFA+}
    \label{fig:mbc_maniAlignLLE_orig}
  \end{subfigure}
  \caption{Cross-Learning}
  \label{fig:x_learn}
\end{figure}
%------------------------------------------------------------------------------------------------------------------
%------------------------------------------------------------------------------------------------------------------
\section{Experimental Results}
\label{sec:sim}
In this section, we validate the proposed concept of cross-learning between SAR and camera images by experimental results. According to our knowledge, a public dataset for concurrent SAR and camera measurements is not available. Therefore, as part of this research, we have carried out such measurements in a laboratory controlled environment. 
Our experimental setup (see Figure~\ref{fig:Ex_setup}) mainly consists of a ZED stereo camera (see \cite{zed} for specifications), a $300$ GHz FMCW radar (see Table \ref{tab:300} for specifications) and a trolley (of size $1 \times 0.5 \times 0.55$ (length $\times$ width $\times$ height) m$^3$) on a turntable.
Figure~\ref{fig:CiS_Msur_schem} shows the measurement schematic of the experiment. 
The trolley is placed on a turntable at a distance of $3.65$ m from the joint sensors (ZED and radar) platform. Measurements from the sensors are taken for every $5$ degree angular turn (counter clock-wise) of the trolley. This emulates the circular motion of the sensors around the target. Thus, $L=72$ (synthetic) aperture samples are obtained all around the trolley.
Note, in this paper, we consider a full $360^\circ$ circular aperture, for the purpose of illustration only. However, in practice, the target can be seen by a partial aperture resulting in corresponding gains via cross-learning.
For every aperture sample, we consider only left lens RGB image from the ZED. We convert the RGB image to a greyscale image. Figure~\ref{fig:greyScale_img} shows the ZED images of the trolley at different positions (on the truntable) $l=1\,(0^\circ), 19\,(90^\circ), 37\,(180^\circ), 55\,(270^\circ)$. 
At every aperture position, the radar scans the target scene for an angular range $\theta = \pm13^\circ$, at angular intervals $\Delta_\theta=0.25^\circ$.
Then, using (\ref{eq:fl}) and (\ref{eq:r_l_img}), a SAR image is created for every $l$th aperture position. Finally, a combined image of the target is achieved by using (\ref{eq:gamma_ij_x}).
Figure~\ref{fig:sar_img} shows the SAR images for $l=1\,(0^\circ), 19\,(90^\circ), 37\,(180^\circ), 55\,(270^\circ)$ and the combined CiS-SAR image of the trolley.
Note, all SAR images have been normalised so that the maximum intensity is unity.
We can see that the SAR images of individual apertures, Figures \ref{fig:sar_1}--\ref{fig:sar_55}, capture viewing-angle dependent information of the target. Nonetheless, the combined image, Figure~\ref{fig:mbc_original}, provides very good imaging result in capturing the complete outline of the target, which re-affirms the enhanced performance of CBP reconstruction algorithm as proposed in \cite{SG_FSSAR_2018}. However, we can see that the handle of the trolley is not very prominent. 

Now, in order to improve SAR imaging results we use the cross-learning concept, employing ML-CCA approach as explained in Section \ref{sect:xlearn}. Using (\ref{eq:pca_coeff_r}) and (\ref{eq:pca_coeff_s}), for $n=m=15$ principal components, we obtain the PCA based manifolds of the images of the two sensors.
Note, we chose these number of principal components as they seemed to provide good results from a qualitative perspective. 
Figure~\ref{fig:pca} shows these manifolds for first two principal components.
We can see that the ZED manifold is more elaborate than the SAR manifold (which is quite concentrated). This shows that the ZED images are more distinguishable than the SAR images. Thus, the SAR images have a big margin of learning from the camera images.
Figure~\ref{fig:mbc_cca} shows the SAR image using ML-CCA approach. 
Note, we essentially use all of the training images as the test images, i.e., $t=1,\cdots, L$, for both the sensors. 
We can see that in comparison to CiS-SAR (Figure~\ref{fig:mbc_original}), ML-CCA image has captured some new information of the target. The walls of the trolley are more prominent. However, the most interesting aspect is the visibility of the trolley handle. Nonetheless, we can see that the target information both in CiS-SAR and ML-CCA does not seem to overlap for every pixel. Thus, a natural course of action is to combine the two images. We name the combined CiS-SAR and ML-CCA image as ML-CCA+. Figure~\ref{fig:mbc_cca_orig} shows the ML-CCA+ image. We can see that it is a much improved image than the CiS-SAR as in Figure~\ref{fig:mbc_original}.

In order to compare the performance of ML-CCA or ML-CCA+, we also provide imaging results with another possible cross-learning approach. In this approach, instead of two levels of abstraction\footnote{Note, here we do not count the manifold creation as a level of abstraction.}, as in ML-CCA, we use a single level of abstraction. It is the manifold alignment (MFA) approach. In this approach the PCA-based manifolds of the two sensors are essentially aligned using Procrustes analysis as in \cite{Wang_ManifoldAlign_2008}.
The basic idea is that given pairwise correspondence between the two datasets (assumed centred), a mapping is obtained to align the test data points. Using the earlier terminology developed in this paper, the following singular value decomposition (SVD) is performed as a first step.
\be
{\rm SVD}(\Pbf_r\Pbf_s^T) = \Ubf \Sigmabf \Vbf^T
\label{eq:evd_r}
\ee
assuming $n=m$.
Then, the improved image via MFA can be obtained as
\be
\tilde{\rbf}_t = k \rbf_t \Qbf
\label{eq:rec_mfa}
\ee
where $k \defeq {\rm trace}(\Sigmabf)/{\rm trace}(\Pbf_r\Pbf_r^T)$ and $\Qbf\defeq \Ubf \Vbf^T$.
Figure~\ref{fig:mbc_maniAlign} shows the imaging result of using MFA approach. Similar to ML-CCA+, we also provide the image result of MFA+ in Figure~\ref{fig:mbc_maniAlign_orig}. We can see that MFA does get some extra information in comparison to CiS-SAR (Figure~\ref{fig:mbc_original}). However, its performance is inferior to ML-CCA (Figure~\ref{fig:mbc_cca}). Similarly, ML-CCA+ (Figure~\ref{fig:mbc_cca_orig}) provides better result than MFA+.

We also compare the performance of ML-CCA or ML-CCA+ with another possible multi-level cross-learning approach. In this approach, we extend the MFA by neighbourhood embedding via LLE (similar to Section \ref{sect:lle}). We name this approach as multi-level MFA (ML-MFA).
Figure~\ref{fig:mbc_maniAlignLLE} shows the imaging result of ML-MFA. Similar to ML-CCA+, we also provide the image result of ML-MFA+ in Figure~\ref{fig:mbc_maniAlignLLE_orig}. We can see that ML-MFA is showing some features of the trolley handle. Therefore, ML-MFA+ shows an improved image. It is better than MFA+ (Figure~\ref{fig:mbc_maniAlign_orig}). However, ML-CCA+ (Figure~\ref{fig:mbc_cca_orig}) still outperforms ML-MFA+. Nonetheless, we can say that multi-level abstraction approaches have superior performance in realising the concept of cross-learning to improve SAR images by using the camera images.
%------------------------------------------------------------------------------------------------------------------
%------------------------------------------------------------------------------------------------------------------
\section{Conclusions}
\label{sec:concl}
In this paper, we have proposed a novel concept of cross-learning, in order to improve SAR images by learning from the camera images.
Despite the fact that the two sensors are very different modalities, 
we have used a multi-level abstraction approach to achieve knowledge transfer between them.
We have shown that a realisation of multi-level abstraction in the form of creation of a coherent subspace over SAR and camera manifolds, followed by neighbourhood embedding, provides very good results.
We have also proposed other possible approaches to materialise the concept of cross-learning, namely, a manifold alignment approach and a multi-level manifold alignment (which includes neighbourhood embedding) approach.
Overall, we have observed that multi-level abstraction approaches provide better performance results. 
In order to validate the proposed concept, we have provided experimental results on real data obtained in controlled lab environment.
Cross-learning is an ongoing research area and this paper highlights some early achievements in this field. 
%------------------------------------------------------------------------------------------------------------------
%------------------------------------------------------------------------------------------------------------------
\bibliographystyle{IEEEtran}

%------------------------------------------------------------------------------------------------------------------
%------------------------------------------------------------------------------------------------------------------
\end{document}